\documentclass[iop]{emulateapj}

\usepackage{graphicx}
\usepackage{amsmath}
\usepackage{subfigure}
\usepackage{lipsum}
\usepackage{float}
\usepackage{color}
\usepackage{array,booktabs}
\usepackage{longtable}
\usepackage{tablefootnote}
\usepackage{footnote}
\usepackage{tabularx}
\usepackage{threeparttable}

\newcounter{ltfootnote}



\newcolumntype{n}{>{\centering\arraybackslash}X}

\newcommand{\teff}{$T_{\rm eff}$}
\newcommand{\logg}{$\log{g}$}
\newcommand{\rearth}{$R_\oplus$}
\newcommand{\kep}{\textit{Kepler}}

\newcommand{\RN}[1]{%
  \textup{\uppercase\expandafter{\romannumeral#1}}%
}

\newcommand{\Teff}{\textnormal{\tiny \textsc{$T_{\rm eff}$}}}
\newcommand{\Logg}{\textnormal{\tiny \textsc{$\log{g}$}}}
\newcommand{\MH}{\textnormal{\tiny \textsc{[M/H]}}}
\newcommand{\mgb}{Mg~{\sc i}~\ensuremath{b}}

\defcitealias{Fressin2013}{F13}
\defcitealias{Brewer2016}{B16}
\defcitealias{Huber2014q16}{H14}
\defcitealias{Buchhave2015}{BL15}
\defcitealias{Wright2012}{W12}
\defcitealias{Dong2014}{D14}
\defcitealias{Johnson2010}{J10}

\slugcomment{Submitted to ApJ on 12/05/2016}

\shorttitle{\kep\ Field Metallicities}

\begin{document}

\title{The metallicity distribution and hot Jupiter rate of the \kep\ field: \\
   Hectochelle High-resolution spectroscopy for 776 \kep\ target stars}

\author{Xueying Guo\altaffilmark{1}, 
John A. Johnson\altaffilmark{2},
Andrew W. Mann\altaffilmark{3},
Adam L. Kraus\altaffilmark{3},
Jason L. Curtis\altaffilmark{4},
David W. Latham\altaffilmark{2}}

\altaffiltext{1}{Department of Physics and Kavli Institute for Astrophysics and Space Research, Massachusetts Institute of Technology, Cambridge, MA 02139, USA}

\altaffiltext{2}{Harvard-Smithsonian Center for Astrophysics, 60 Garden Street, Cambridge, MA 02138, USA}

\altaffiltext{3}{Department of Astronomy, The University of Texas at Austin, Austin, TX 78712, USA}

\altaffiltext{4}{Department of Astronomy, Columbia University, 550 West 120th Street, New York, New York 10027, USA}

\email{shryguo@mit.edu}

\begin{abstract}
\setlength{\parindent}{2ex}

The occurrence rate of hot Jupiters from the \kep\ transit survey is roughly half that of radial velocity surveys targeting solar neighborhood stars. One hypothesis to explain this difference is that the two surveys target stars with different stellar metallicity distributions. To test this hypothesis, we measure the metallicity distribution of the \kep\ targets using the Hectochelle multi-fiber, high-resolution spectrograph. 
Limiting our spectroscopic analysis to 610 dwarf stars in our sample with \logg\ $> 3.5$, we measure a metallicity distribution characterized by a mean of [M/H]$_{\rm mean} = -0.045 \pm 0.009$, in agreement with previous studies of the \kep\ field target stars. In comparison, the metallicity distribution of the California Planet Search radial velocity sample has a mean of [M/H]$_{\rm CPS, mean} = -0.005 \pm 0.006$, and the samples come from different parent populations according to a Kolmogorov-Smirnov test. We refit the exponential relation between the fraction of stars hosting a close-in giant planet and the host star metallicity using a sample of dwarf stars from the California Planet Search with updated metallicities. The best-fit relation tells us that the difference in metallicity between the two samples is insufficient to explain the discrepant Hot Jupiter occurrence rates; the metallicity difference would need to be $\simeq$0.2-0.3 dex for perfect agreement. We also show that (sub)giant contamination in the \kep\ sample cannot reconcile the two occurrence calculations. We conclude that other factors, such as binary contamination and imperfect stellar properties, must also be at play.

\end{abstract}


\section{Introduction}

The primary \kep\ mission observed $\approx 2\times 10^{5}$ target stars during its four-year lifetime. As of June 2016, 4696 planet candidates have been identified \citep{Coughlin2016}, among which 2290 planets have been confirmed or statistically validated \citep[e.g.,][]{Morton2016}.
This large dataset is a powerful tool for exoplanet statistics, and has therefore spawned a number of planet occurrence studies.
\citet{Howard2012} reported the occurrence rate as a function of planet radius, orbital period, and host star effective temperature for all \kep\ planet candidates with periods less than 50 days. \citet{Fressin2013} (hereafter F13) simulated the \kep\ mission based on the observations from Q1--Q6, estimated the false positive probabilities, and calculated the occurrence rate for planet of different sizes and orbital periods. \citet{Petigura2013} studied the prevalence of earth-size planets orbiting sun-like stars. \citet{Dressing2015} and \citet{Gaidos2016} improved the estimation of occurrence of planets orbiting M dwarfs. \citet{Foreman2014} presented a general hierarchical probabilistic framework to analyze the exoplanet populations and measured the rate density of Earth analogs. These and many other studies on the planet occurrence and distribution in parameter space provide vital observational tests for theories of planet formation and migration.

All previous occurrence rate studies have reported that small planets ($R_p \lesssim 2$~\rearth) are the most abundant; however, since giant planets are the easiest to be detected with radial velocity (RV) and transit methods, their measured occurrence are less subject to complications that grow with decreasing S/N, like pipeline completeness \citep{Christiansen2016}. Giant planets with periods $P<10$ days and $M_p > 0.1\rm M_{\rm Jupiter}$ are dubbed as hot Jupiters (HJ)\footnote{Definitions of HJs in other works only have minor differences in criteria} in \citet{Wright2012} (hereafter W12), and much effort has been devoted into the study of the HJ occurrence rates both in the \kep\ field and the solar neighborhood. 

It has been established that there is a discrepancy between the HJ occurrence rates of the \kep\ survey and that of the RV survey of solar neighborhood. \citet{Howard2012} estimated a $0.4\pm 0.1\%$ occurrence for planets around G/K dwarfs in the \kep\ survey with periods shorter than 10 days and radius between $8R_{\oplus}$ and $20R_{\oplus}$, and \citetalias{Fressin2013} reported a $0.43\pm0.05\%$ occurrence rate for planets in the \kep\ survey with periods between 0.8 days and 10 days and radius between $6R_{\oplus}$ and $8R_{\oplus}$. These are in agreement with HJ occurrence measurements from other transit surveys, e.g., $0.31^{+0.43}_{-0.18}$\% from the OGLE-III Survey \citep{Gould2006} and $0.10^{+0.27}_{-0.08}$\% from the SuperLupus Survey \citep{Bayliss2011}. RV surveys have found HJ occurrence rates a factor of two higher than those from \kep\ and similar transit surveys. \citetalias{Wright2012} analyzed the whole sample of California Planet Search (CPS) and found that the HJ occurrence rate is $1.20\pm0.38\%$, consistent with $1.2\pm0.2\%$ from \citet{Marcy2005} and $0.89\pm0.36$\% from \citet{Mayor2011}. A comparison between the occurrence of any individual RV survey and the results from \kep\ are only marginally significant, but the consistency between independent RV studies makes the discrepancy highly significant.  

\citetalias{Wright2012} proposed that a possible reason for this discrepancy is the metallicity difference between the transit targets and the RV targets, since it has been found by radial velocity surveys that intrinsic probability for a dwarf star to host a giant planet depends on the metallicity and the mass of the host star \citep{Santos2004, Fischer2005, Bowler2009,Johnson2010}, and transit survey targets are slightly above the galactic plane in general, thus may be more metal poor than RV targets. \citet{Johnson2010}(hereafter J10) established an exponential relation between the probability of a star harboring a giant planet and its metallicity: $f\propto 10^{1.2\rm [Fe/H]}$; \citet{Fischer2005} reported a similar relation with a different exponential index: $P(\rm planet)=0.03\times10^{2.0\rm [Fe/H]}$. Since \kep\ stars are located further away in a different region of the Galaxy, and the \kep\ field is slightly tilted above the galactic plane, \kep\ stars are in general further away from the Galactic plane than stars in the solar neighborhood, and thus it is reasonable to wonder whether the \kep\ target stars are slightly more metal poor than the CPS stars in the solar neighborhood. To test this notion, we need to obtain reliable measurements of metallicity distributions for both the \kep\ sample and the RV sample. Apart from the metallicity difference, other factors that may contribute to the HJ rate difference will be discussed in section \ref{section6}.

Precise metallicities can be measured by fitting high-resolution spectra over a wide wavelength range. \citet{Buchhave2012} introduced a tool, Stellar Parameter Classification (SPC), that accurately measures stellar metallicity using a grid of library spectra with varying stellar parameters to match the observed spectra originating from different instruments, and determines values of \teff, \logg\ and [M/H] for stars by finding out the best-match parameters assuming that the cross-correlation function peak heights vary smoothly between grid points. 

\citet{Valenti1996} described the software package, Spectroscopy Made Easy (SME), that fits stellar spectra by calculating model atmospheres and adjusting free parameters, which is computationally costly but copes with the errors from coarse grid interpolations well. MOOG is another widely used package to perform spectral analysis, which outputs the chemical composition of a star by fitting absorption lines with model atmosphere, assuming local thermodynamic equilibrium \citep{Sneden1973}. In addition, \citet{Petigura2015} introduced a new stellar characterization tool, SpecMatch, which is designed to study faint \kep\ stars and extracts various stellar properties from high-resolution optical spectra by matching the observed spectra with the synthetic stellar spectra library from \citet{Coelho2005}. 

\citet{Valenti2005} measured metallicities for 1024 stars from CPS using SME on stellar spectra observed by Lick and Keck Observatories, and the Anglo-Australia Telescope. \citet{Brewer2016} (hereafter B16) developed a semi-automated procedure to fit the line parameters using SME and provided an extended abundance analysis for 1626 CPS stars. The size of the \kep\ target sample makes it impractical to observe high-resolution spectra and measure the precise metallicity for every star, but it is possible to measure the sample's overall metallicity distribution from a large subsample. \citet{Dong2014} (hereafter D14) reported a mean iron abundance $\rm [Fe/H]_{mean} = -0.04$ for the \kep\ sample by measuring iron abundances for 12,000 \kep\ host stars with the Large Sky Area Multi-Object Fiber Spectroscopic Telescope (LAMOST) low-resolution spectroscopic survey data. 

Here we measure the metallicity distribution of \kep\ target stars through a sample of 835 high-resolution spectra taken with the Hectochelle multi-fiber spectrograph. We develop a functional form to fit the time-dependent and fiber-dependent Hectochelle continuum profile simultaneously with spectral line profiles using the calibrated Kurucz synthetic library \citep{Kurucz1970}, and interpolate the model grids to find out the best-fit stellar parameters. In the end, we obtain metallicities for 776 \kep\ stars which we use to measure the overall metallicity distribution of dwarf stars for the \kep\ survey.

We describe our sample construction and observations in Section \ref{section2}, and present the functional formula for the Hectochelle continuum as well as the detailed procedure of fitting with the Kurucz library and measuring stellar parameters in Section \ref{section3}. In Section \ref{section4}, we show that our analysis reproduces the solar properties with a set of Hectochelle twilight spectra, stellar properties for 27 stars with overlapping Hectochelle and other high-resolution spectra, and the established metallicity of NGC 752 when applied to member stars. In Section \ref{section5} we present the metallicities of 776 \kep\ stars, report the dwarf star metallicity distribution of the \kep\ survey, and compare with the LAMOST distribution and the CPS star metallicity distribution. Finally, we summarize our work and discuss the implications of the reported metallicity distribution on the hot Jupiter occurrence rate and possible directions of future studies in Section \ref{section6}.

\begin{figure}
  \includegraphics[width=\linewidth]{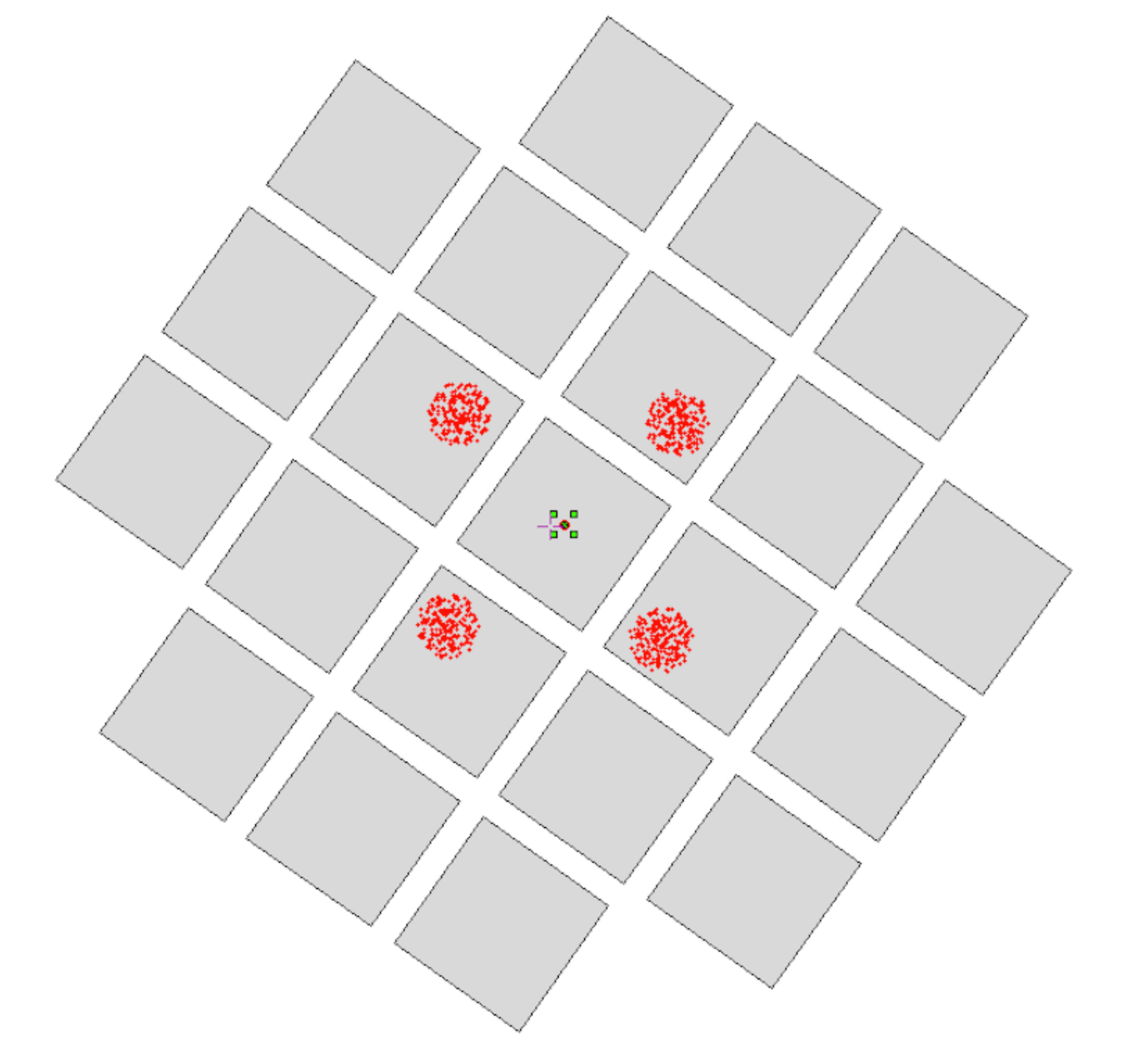}
  \caption{\kep\ has 21 CCD modules, each of which covers a 5 square degree field of view, as is shown with the gray squares in the figure. Positions of all stars in our sample are marked with red dots on top of the \kep\ field.} 
  \label{fig:FoV}
  \bigskip
\end{figure}

\bigskip

\section{Observation and Sample} \label{section2}

\subsection{Hectochelle Observations }

Hectochelle is a high-resolution ($R \approx 34,000$) single-order, multi-object echelle spectrograph \citep{Hecto_observ_guide}. It has 240 fibers which can be deployed in a one square degree field, 20 of which were reserved for sampling the sky. We filled as many of the remaining 220 fibers with selected \kep\ target stars as possible given fiber allocation limitations. Four observations pointing at four different fields within the \kep\ field were obtained. Observed fields were selected to be evenly distributed around the central area of the \kep\ field to get a representative sampling of target stars, which we show in Figure \ref{fig:FoV}. Two of the pointings had exposure times of 900 seconds, and the other two had exposure times of 600 seconds. All stellar spectra were taken through Hectochelle's RV31 filter, which covers a wavelength range from 5145 to 5300~\AA, and contains the gravity-sensitive \mgb\ lines. Relevant information about the four observations is shown in Table \ref{tab:pointings}.

\begin{table*}
\renewcommand*{\arraystretch}{1.4}
    \caption{Hectochelle Observations}
    \centering
    \begin{tabularx}{\textwidth}{nnnnnnnnn}
    \hline\hline
    Pointing ID & Central RA(J2000) & Central Dec(J2000) & Epoch & Exposure Time(seconds) & Filter ID & Resolution(\AA) & Number of Stars \\
    \hline
    Kep07\_1 & 19h14m17.41s & +42d42m07.58s & 2014-07-15 & 900 & RV31 & 0.034 & 210 \\
    Kep08\_1 & 19h31m39.00s & +46d09m40.81s & 2014-07-15 & 600 & RV31 & 0.034 & 206 \\
    Kepbinary1\_1 & 19h32m17.93s & +42d54m37.90s & 2014-07-15 & 900 & RV31 & 0.034 & 210 \\
    Kepbinary2\_1 & 19h12m21.56s & +42d01m32.68s & 2014-07-15 & 600 & RV31 & 0.034 & 209 \\
    \hline
    \end{tabularx}
    \label{tab:pointings}
\bigskip
\end{table*}

\subsection{Sample Description}
To get high-quality spectra and best emulate the observed \kep\ stellar population, we primarily observed stars with $K_p < 15$ considering brightness limitations. To remove targets selected exclusively for guest observer (GO) programs we only included targets with at least two quarters of \kep\ data. Pointings were selected to give a range of Galactic coordinates representative of the \kep\ sample. For each pointing, targets were fed into the Hectochelle target selection tool {\it xfitfibs}, which selects targets based on guide star and fiber positioning restrictions as well as user-provided priorities. Since we provided no priorities the selection was nearly random. In the end, we obtained spectra for 835 stars.
Excluding 21 spectra with extremely low signal to noise ratio per pixel ($SNR$ hereafter), 22 spectra that suggest high temperatures or very fast rotation, and 16 spectra indicative of very low temperatures (deep and wide absorption lines), we obtained a final sample of 776 stars.

\begin{figure}
 \includegraphics[width=\linewidth]{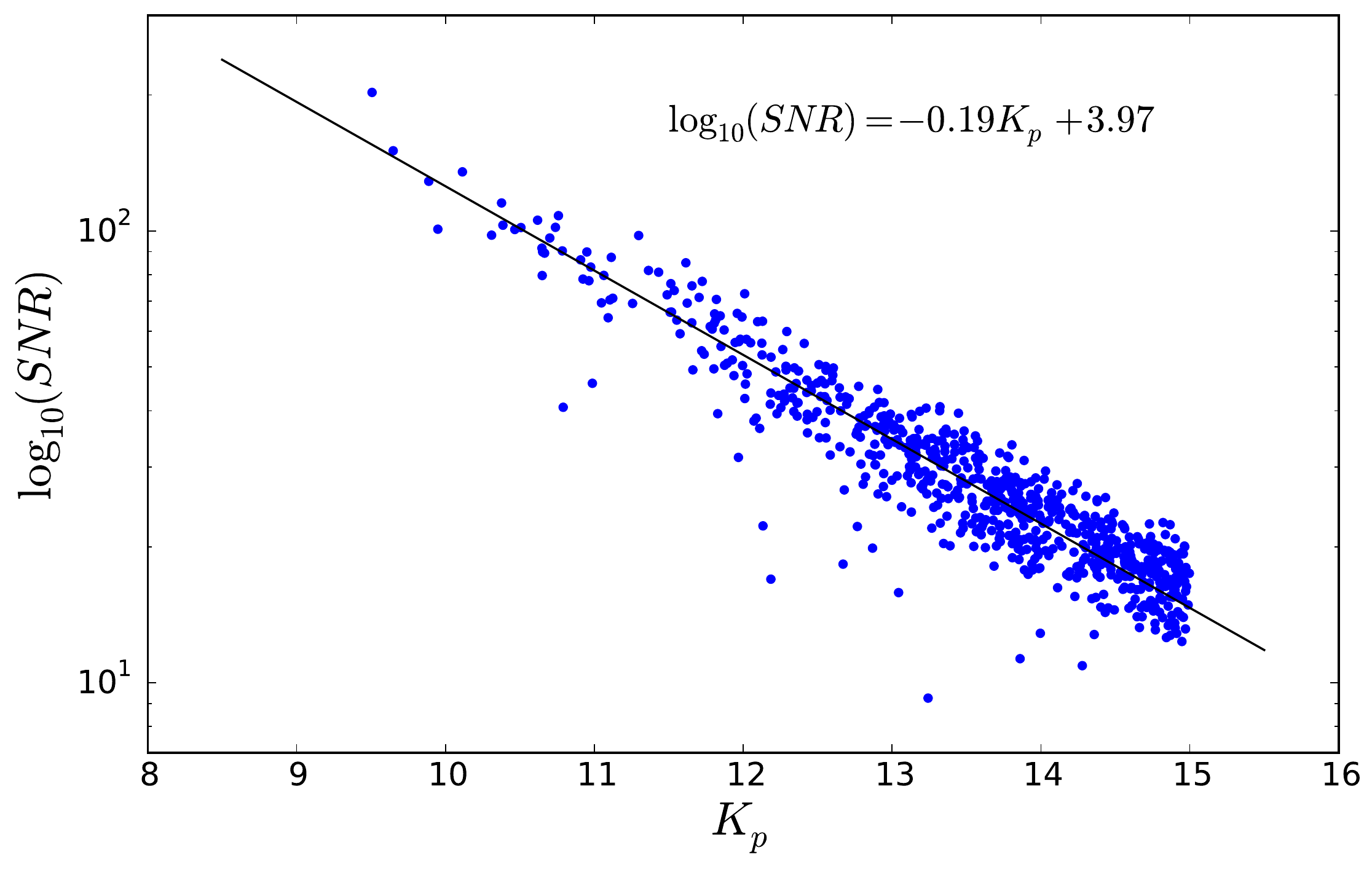}
  \caption{Relation between $\rm log_{10}$($SNR$) and \kep\ magnitude of stars in our sample. The $SNR$ of each star was calculated by taking the square root of the average of the highest 2\% flux values in its spectrum. Half of our sample were observed with 900 seconds exposure and the other half observed with 600 seconds exposure and applied a $\sqrt{3/2}$ factor to emulate the 900 seconds exposure when calculating their $SNR$s. Blue dots in the figure show data for individual stars, and the black curve represents the best fit relation, see equation (\ref{eq:first}).}
  \label{fig:SNR_Kepmag}
  \bigskip
\end{figure}

We estimated the SNR for each spectra using $SNR=\sqrt{f_{ave}}$, where $f_{ave}$ is the average continuum flux level, and fit an linear relation between $\rm log_{10}$($SNR$) and \kep\ magnitude. A $\sqrt{3/2}$ factor is applied to stars with 600 seconds exposure time, so that all $SNR$s used in the fit represent Hectochelle observations with 900 seconds exposure time. The $\rm log_{10}$($SNR$) versus \kep\ magnitude relation for our sample is shown in Figure \ref{fig:SNR_Kepmag}, which could be used as an $SNR$ reference for future Hectochelle observations. The best fit empirical relation is given by
\begin{equation} \label{eq:first}
\begin{aligned}
\mathrm{log_{10}}(SNR) = -0.19 K_p+3.97
\end{aligned}
\end{equation}
\noindent where $K_p$ is the \kep\ magnitude of a given target star.

\begin{figure}
  \includegraphics[width=\linewidth]{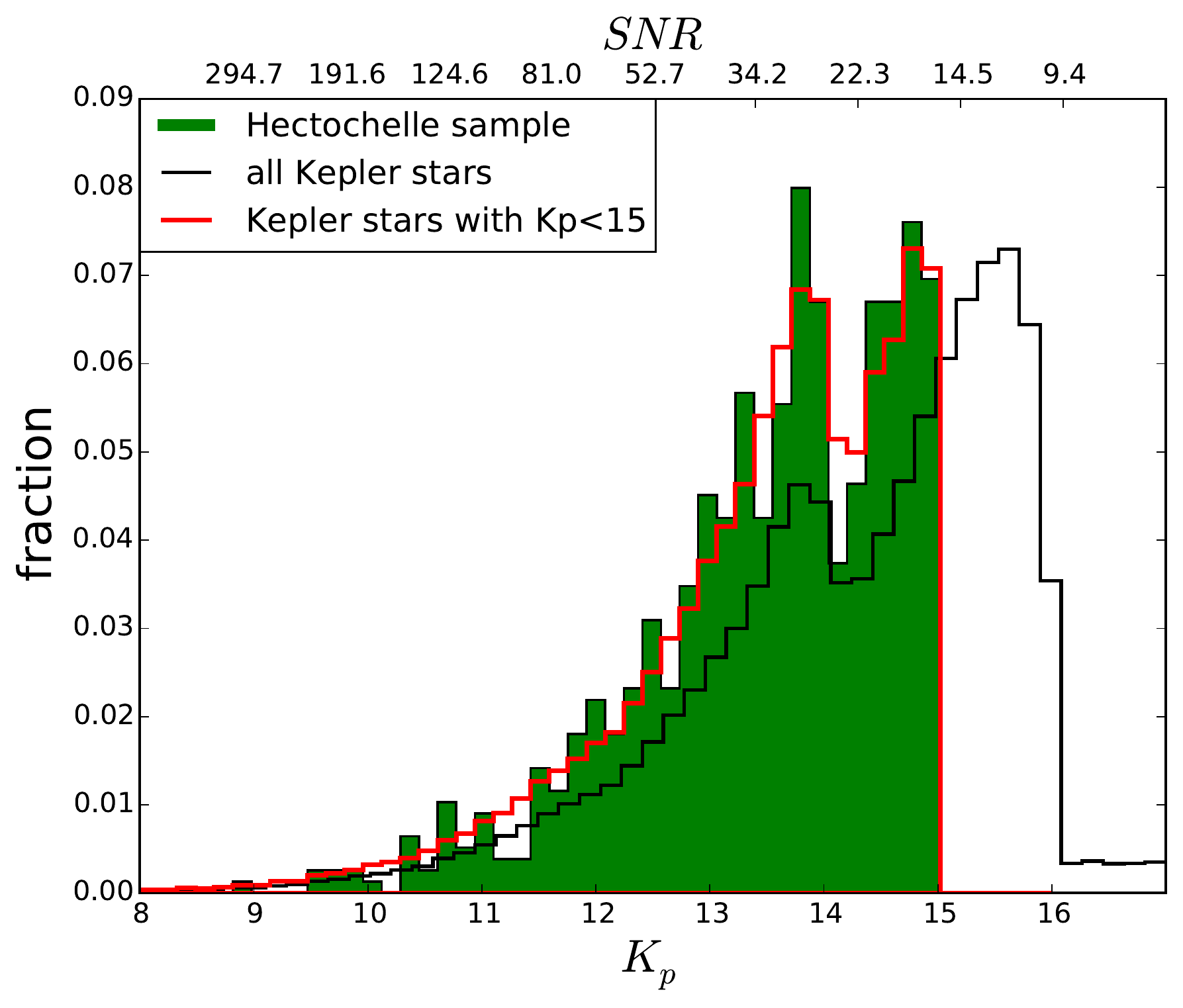}
  \caption{$K_p$ distribution of our final sample of 776 stars, which all have $K_p$\textgreater 15, is shown with the green histogram. The black histogram edge shows the distribution of the whole \kep\ target star sample, and the red histogram edge represents the distribution of \kep\ stars with $K_p<15$. We also label the $SNR$ corresponding to each \kep\ magnitude on top, based on the empirical relation between $SNR$ and $K_p$ shown with equation (\ref{eq:first}).}
  \label{fig:Kepmag_distribution}
  \bigskip
\end{figure}

The $K_p$ distribution of our final sample is shown as the green histogram in Figure \ref{fig:Kepmag_distribution}. The black histogram edge shows the distribution of the whole \kep\ target star sample, and the red histogram edge represents the distribution of \kep\ stars with $K_p<15$. We also put a $SNR$ axis in Figure \ref{fig:Kepmag_distribution} according to the $SNR-K_p$ relation shown with equation (\ref{eq:first}), thus indicating the approximate $SNR$ distribution for our sample. All stars in our final sample have $SNR$\textgreater 15, and about 68\% stars have $SNR$\textgreater 20. In addition, the $K_p$ distribution of our sample have approximately the same shape as that of the entire \kep\ sample when applied a cut at $K_p=15$.

\begin{figure} 
  \includegraphics[width=\linewidth]{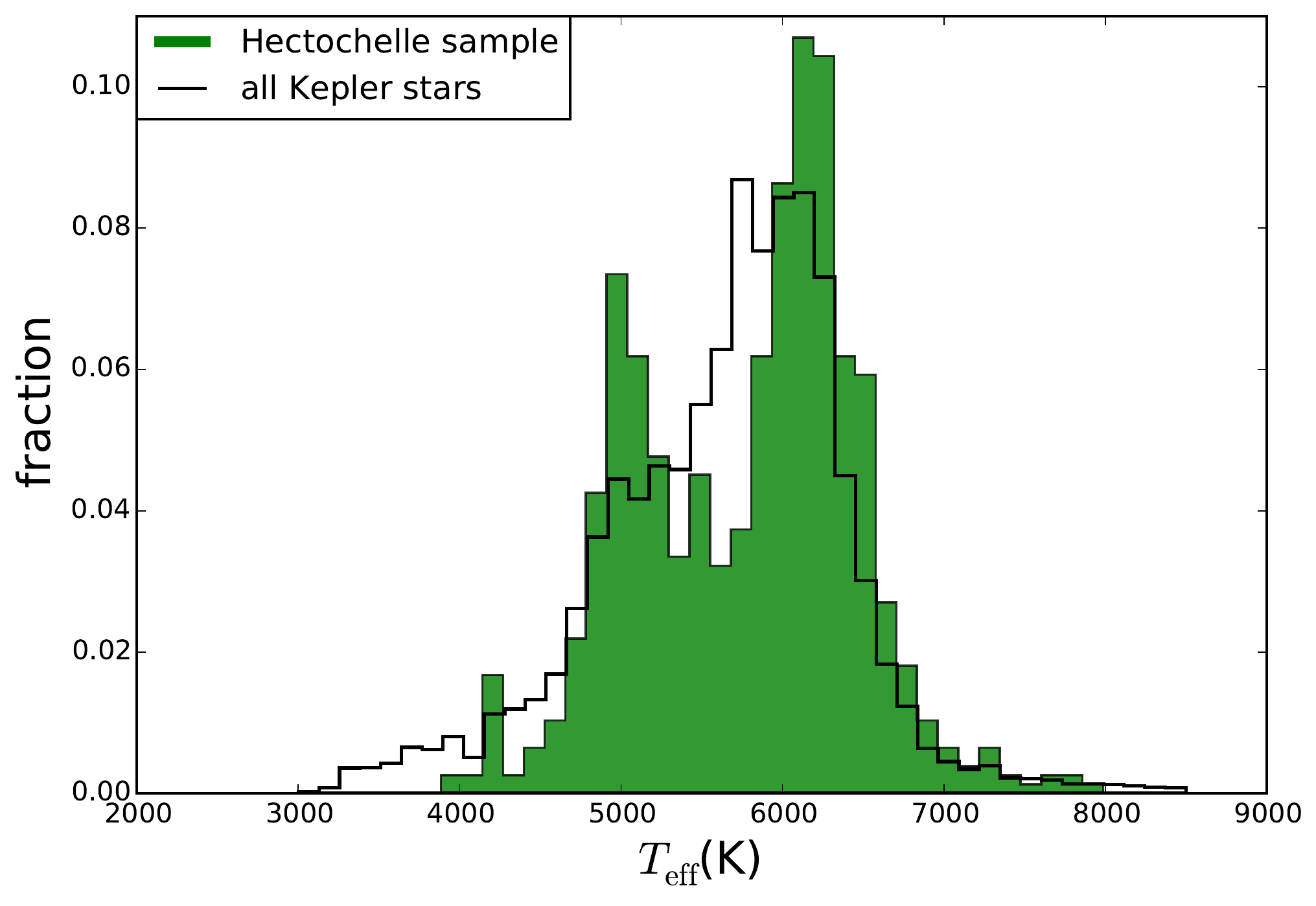}
  \caption{\teff\ distribution of our final sample of 776 stars. We notice that the stars concentrate around 5000K and 6000K, which is consistent with the \teff\ distribution of the whole \kep\ sample as shown with the black line, and studied in \citetalias{Huber2014q16}.}
  \label{fig:Teff_q17_distrib}
  \bigskip
\end{figure}

The effective temperature (\teff) distribution of the sample is shown in Figure \ref{fig:Teff_q17_distrib}. The \teff\ values are from the \kep\ Q1-Q17 Stellar Parameters Archive (available at the Mikulski Archive for Space Telescopes), which was compiled and updated by \citet{Huber2014q16} (hereafter H14). The \kep\ Q1-Q17 Stellar Parameters Archive (hereafter KSPA) comprises mostly (around 70\%) \kep\ Input Catalog (KIC) photometry results, supplemented by asteroseismology and spectroscopy measurements from literature, so the systematic offsets in \logg\ measurements \citep{Verner2011, Everett2013} and the large uncertainties in [Fe/H] measurements from the KIC are reflected in the KSPA \citep{Huber2014q16}. Although there are systematic biases in stellar temperatures from different methods, the offsets are usually $\lesssim$ 200K. Therefore we used \teff\ values from the KSPA as a reference, and found that $\simeq$99\% of the 776 stars in our final sample are FGK stars with a \teff\ range from 3700~K to 7500~K, and $\simeq$89\% stars in our final sample have \teff\ in the range from 4200~K to 6500~K.

There are 2 peaks in the \teff\ distribution, one around 5000K and the other around 6000K. This is similar to the \teff\ distribution of the whole \kep\ target stars sample. Although the shapes of the two distributions are not identical, it is likely that our observed sample has a similar \teff\ distribution to that of the parent \kep\ sample.
Moreover, the two stellar population peaks around 5000~K and 6000~K, indicated in Figures 9 and 12 of \citetalias{Huber2014q16}, are consistent where we find population peaks; around 6000~K for the dwarfs, and around 5000~K corresponding to a population peak of subgiants and giants. This gives us an opportunity to investigate the level of subgiant/giant contamination in the \kep\ sample \citep{Mann2012, Wang2015}, as well as its effects on the calculation of HJ occurrence rate, which we discussed in Section \ref{section5}. 

\bigskip

\section{Spectroscopic Analysis} \label{section3}

The spectra of all 835 stars were extracted from the echelle frames using the standard Hectochelle data reduction pipeline. Cosmic ray flares were subtracted, and 10~\AA\ on each end of the wavelength range was cut off to avoid instrumental distortion, thus still leaving us with a wavelength range from 5155~\AA\ to 5290~\AA. 

Unlike typical spectrographs whose continuum can be described by a simple blaze function or removed by a polynomial fitting or well-calibrated by the combination of different orders \citep[e.g., HIRES,][]{Becker2013}, Hectochelle's instrumental continuum profile has a complicated sinusoidal shape as is shown in the central panel of Figure \ref{fig:demo_contin}. The shape is highly fiber- and time-dependent. As a result, the continuum removal is a significant problem preventing Hectochelle from being used for precise spectroscopic analysis until now. In this work, we constructed a functional form with which the Hectochelle continuum is fitted. Model construction and tests are described below.

\subsection{Continuum Model} \label{bozomath}

\begin{figure*}
\includegraphics[width=\linewidth]{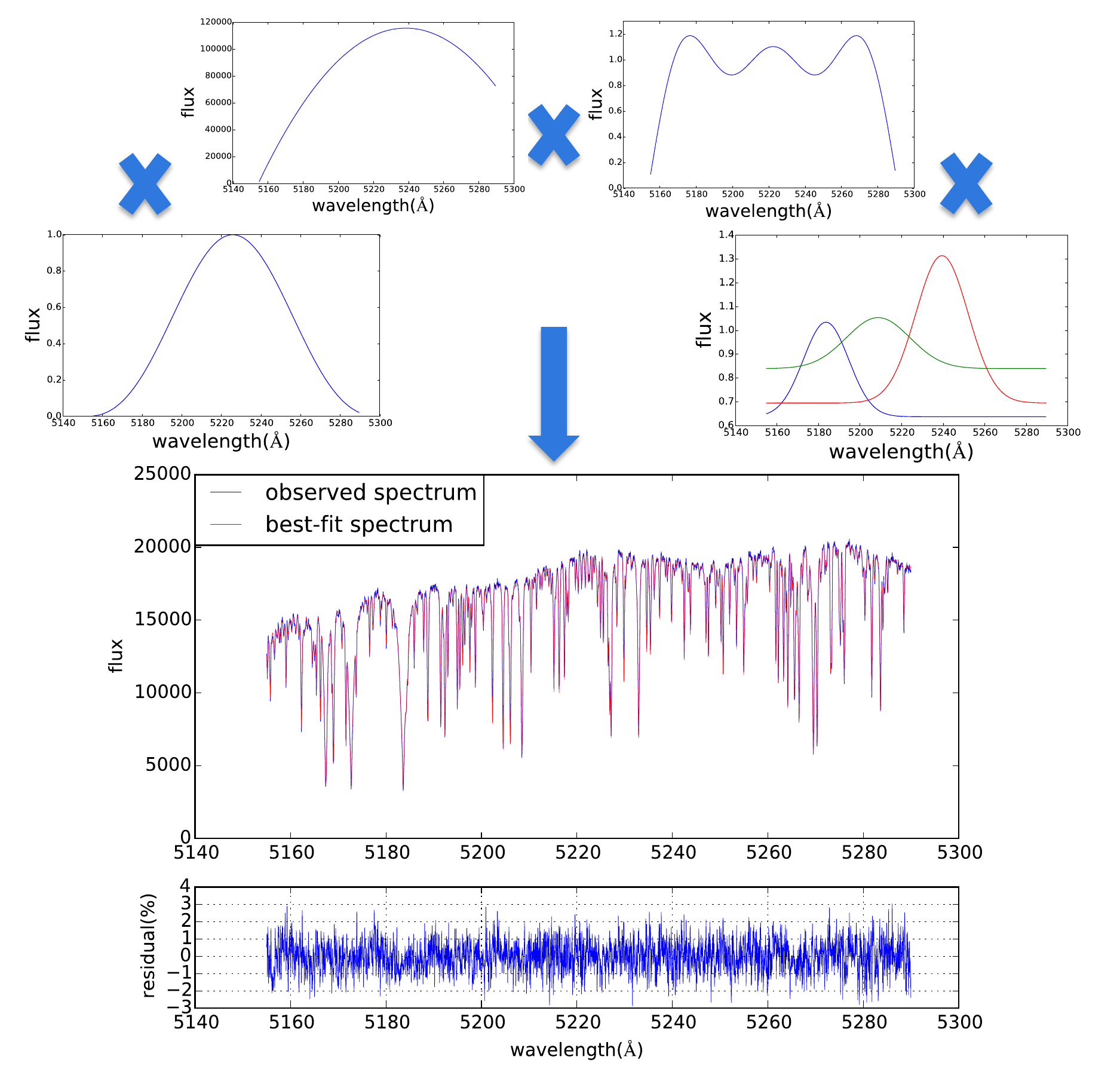}
\caption{The four upper plots show the components of our continuum functional form, which consists of the square of a sinc function to emulate the blaze function, a quadratic, the first 3 terms of a square wave function to emulate the filter transmission function and 3 freely varying Gaussian shapes to make sure there is no structure left in the continuum. The lower plot shows the Hectochelle twilight spectrum in comparison with NSO solar spectrum applied with the corresponding best-fit continuum and the line property parameters (red curve). On the bottom of the plot is the residual between Hectochelle twilight spectrum and the best-fit NSO solar spectra, which is on the level of 2\% throughout the wavelength range.}
\label{fig:demo_contin}
\bigskip
\end{figure*}

We tested various functional forms for the Hectochelle continuum by applying them on the normalized standard high-resolution ($\Delta \lambda \approx$0.006\AA) solar spectrum taken by the National Solar Observatory (NSO) from Kitt Peak, and comparing the output spectrum with a standard twilight spectrum taken by Hectochelle with 30 seconds exposure time and a $SNR$ of around 120 (see Figure \ref{fig:demo_contin}). 

For each parameterized functional form, we multiplied it by the normalized NSO solar spectrum, using a free parameter to describe the line shift. We then convolved it with a Gaussian function with a free parameter to account for the line broadening difference between the instruments. In addition, we added a freely varying constant background level. Lastly, we re-sampled the output NSO solar spectrum onto the Hectochelle wavelength scale, so that the two spectra could be compared by calculating the flux difference. By running an MCMC fitting procedure targeting on minimizing the flux array differences with free parameters representing the continuum profile and the line properties all varying, we obtained a set of best-fit parameters. To determine the effectiveness of the continuum functional form we examined the minimized flux differences as a function of wavelength.

Inspired by previous work on spectrograph continuum removal \citep{Becker2013}, we first tried a polynomial fit. Since the Hectochelle continuum shape has 5 extrema (see Figure \ref{fig:demo_contin}), the polynomial has to be at least 6th-order, which makes it difficult to set priors on the parameters, leading to a long burn-in time during the fit and making it impractical for our work which deals with a large sample and includes multiple model fits for each star in the sample. Such high-order polynomial also suffers from severe oscillations within the data gaps \citep[Runge's phenomenon,][]{Runge1901}, which in our case is the region with wide deep absorption lines that are masked out during continuum fitting. 
Lastly, the high-order polynomial is sensitive to small changes in data values or parameter values. We tried a spline fit across the highest 1\% flux data of each spectrum to emulate the continuum. However, the lack of data in the absorption region made the spline not smooth enough and the continuum level usually underestimated. 

Thus, instead of using a polynomial or spline fit, we took advantage of the fact that the continuum shape consists of a distorted blaze function, a filter transmission function, and an arbitrary instrumental shape residual. We emulated the distorted blaze function with the square of a sinc function multiplied by a quadratic, and built the filter transmission function with the first three terms of a square wave function. In addition, we complemented this basic setup with three freely varying Gaussian shapes to make sure no continuum structure was left in the normalized spectra. A demonstration of the continuum components is shown in Figure \ref{fig:demo_contin}. The final functional form of the continuum is:

\begin{equation}\label{eq:second}
f(x) = \prod_{i=1,2,3,4} f_i(x),
\end{equation}

\noindent
where
\begin{equation}
%\begin{align*}
f_1(x) = \left[\sin\left(2\pi p_0\left(x-p_1\right)\right)/\left(2\pi p_0\left(x-p_1\right)\right)\right]^2, \\
\end{equation}
\begin{equation}
f_2(x) = \left(x-p_2\right)^2 + p_3, \\
\end{equation}
\begin{equation}
f_3(x) = \dfrac{4}{\pi} \left[ \sum_{j=4,5,6} p_j \cdot \sin\left(\left(2j-7\right) p_0 \pi \left(x-p_1\right)\right) \right], \\
\end{equation}
\begin{equation}
f_4(x) = \prod_{k=8,11,14} \left(\dfrac{1}{\sqrt{2\pi}p_k} e^{-\left(x-p_{k-1}\right)^2/2p_k^2} + p_{k+1}\right),
%\end{align*}
\end{equation}
\noindent
where $f(x)$ is the continuum flux as a function of wavelength x, $f_i(x)$ are 4 components of the continuum function, and $p_j$ the 16 fitting parameters. We used 3 additional parameters for line broadening, line shift and constant background level respectively, so the fitting procedure included 19 parameters in total. 

The lower plot of Figure \ref{fig:demo_contin} shows the comparison between a Hectochelle solar spectrum and the corresponding fitted NSO solar spectrum. The residual is around 2\% throughout the whole wavelength range and there is no obvious structure left in the residual, indicating that our continuum removal is effective. The same continuum functional form and fitting procedure were applied on solar spectra taken from the rest of the Hectochelle fibers, and there was no structure shown in the flux residual distribution from any fibers. In addition, although the instrumental profile may appear different for different Hectochelle observations, the same functional form still successfully reproduce the continuum profiles from different observations in our tests in Section \ref{section4}.

\subsection{Measuring Stellar Parameters} 

For each observed stellar spectrum, stellar parameters were obtained by fitting with a subset of the calibrated Kurucz synthetic library combined with our continuum model, using a figure-of-merit for each (\teff, \logg, [M/H]) 3D space grid point, and then interpolating the figure-of-merit distribution in the (\teff, \logg, [M/H]) space and searching for the minimum. The synthetic library was calculated with the Kurucz stellar models \citep{Kurucz1970}, and the grid of the library spans a \teff\ range of 3500--9750 K at increments of 250 K, a \logg\ range of 0.0--5.0 with a step size of 0.5, and a [M/H] range of $-2.5$--0.5 in steps of 0.5 dex. For each star, we took a cuboid subset of the library centered on the \teff, \logg, and [M/H] given by the KSPA, with 3 side lengths being $\Delta$\teff=1500K, $\Delta$\logg= 3.0 and $\Delta$[M/H]= 2.5 respectively. We assumed that this cuboid should contain the grid point corresponding to the actual stellar parameters of the star of interest. The detailed procedure is described below. 

To speed up the fitting with each library spectrum, we first tried to obtain an initial guess of the continuum parameter set for every star. To do this, ``non-absorption regions'' of the spectra were extracted following \citet{Pineda2013}. Specifically, we cut the whole 135~\AA\ spectra into 4~\AA\ wide chunks, and took data points in each chunk which satisfy $f>f_{max}-0.01\times \left(f_{max}-f_{min}\right)$, where $f$ represents each flux value, and $f_{max}$ and $f_{min}$ are the highest and the lowest flux in that chunk. With this criterion, we were able to achieve a balance between keeping a sufficient number of data points to construct an initial continuum model and excluding most of the absorption structures. The extracted ``continuum points'' were fitted with the form defined with equation (\ref{eq:second}), and the resulting parameters are set as the initial guess of the continuum when fitting with each library spectrum. 

Using this continuum as the start point, we initialized the Levenberg--Marquardt minimization algorithm, and obtained the best-fit 19 parameters for each library spectrum. During the fit, library wavelength arrays were re-sampled to match the observed wavelength array. The figure-of-merit of each library spectrum is defined as the relative distance between the observed flux array and best-fit library flux array: 
\begin{equation*}
\delta \equiv \left[\dfrac{\sum_{i=1}^{N}\left(f_{i,obs}-f_{i,best-fit}\right)^2} {N} \right]^{1/2} \dfrac{1}{f_{average}},
\end{equation*}
 
\noindent
where $\delta$ is the figure-of-merit, $f_{i,obs}$ and $f_{i,best-fit}$ are observed flux and best-fit flux on the $ith$ wavelength point, $N$ is the total number of data points in the spectrum, and $f_{average}$ is the average flux level of the observed spectrum. Assuming that the observed spectrum has the same noise level at each wavelength, we have $\delta \propto -\ln{\mathcal{L}}$, where $\mathcal{L}$ is the likelihood of the fitted spectrum being the true spectrum of the star given the observed stellar spectrum, so minimizing $\delta$ is equivalent to maximizing the likelihood. 

\begin{figure*}
\begin{center}
		\subfigure
		{%
		\label{fig:first}
		\includegraphics[width=0.45\textwidth]{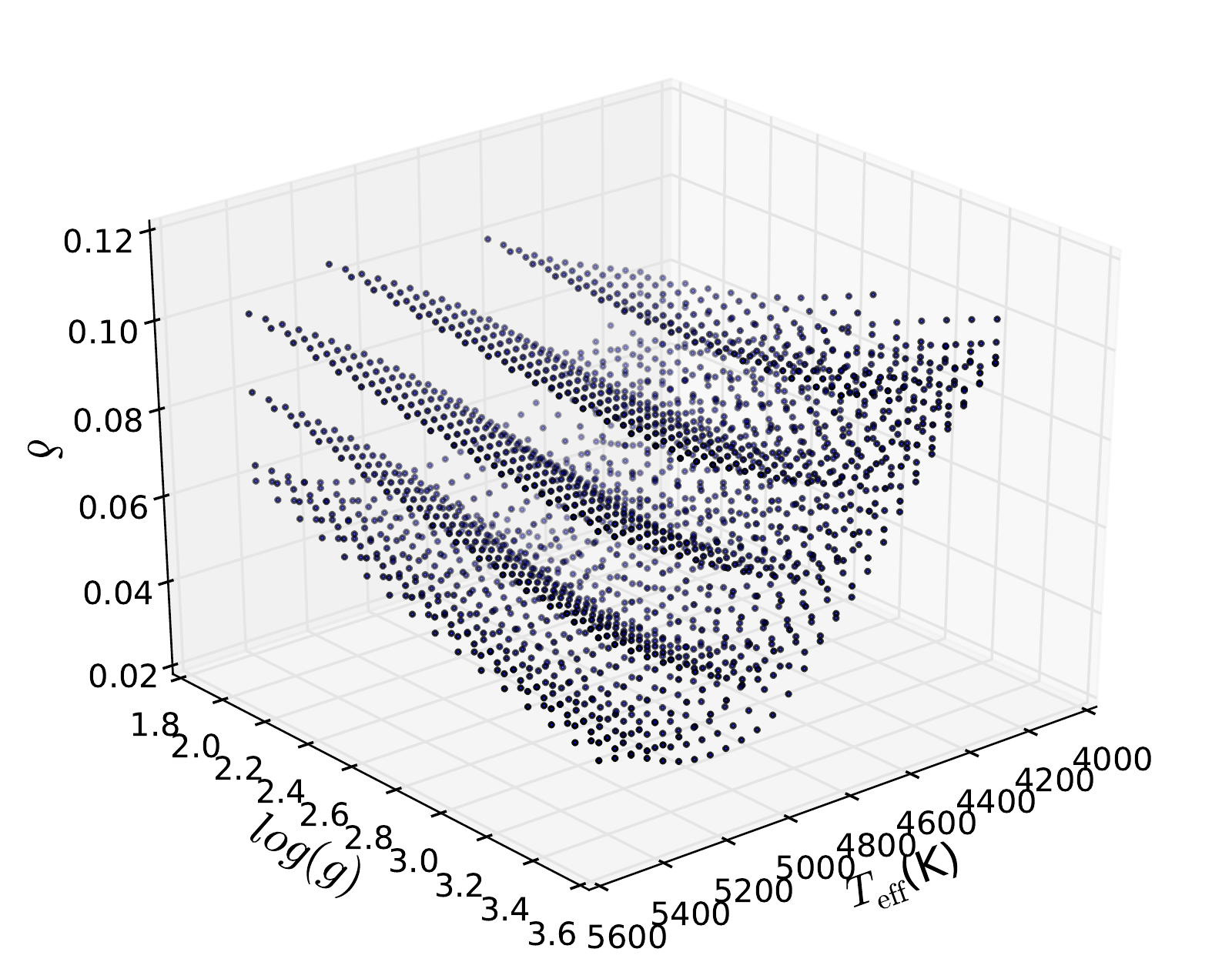}
		}%
        		\subfigure
		{%
           	\label{fig:second}
           	\includegraphics[width=0.45\textwidth]{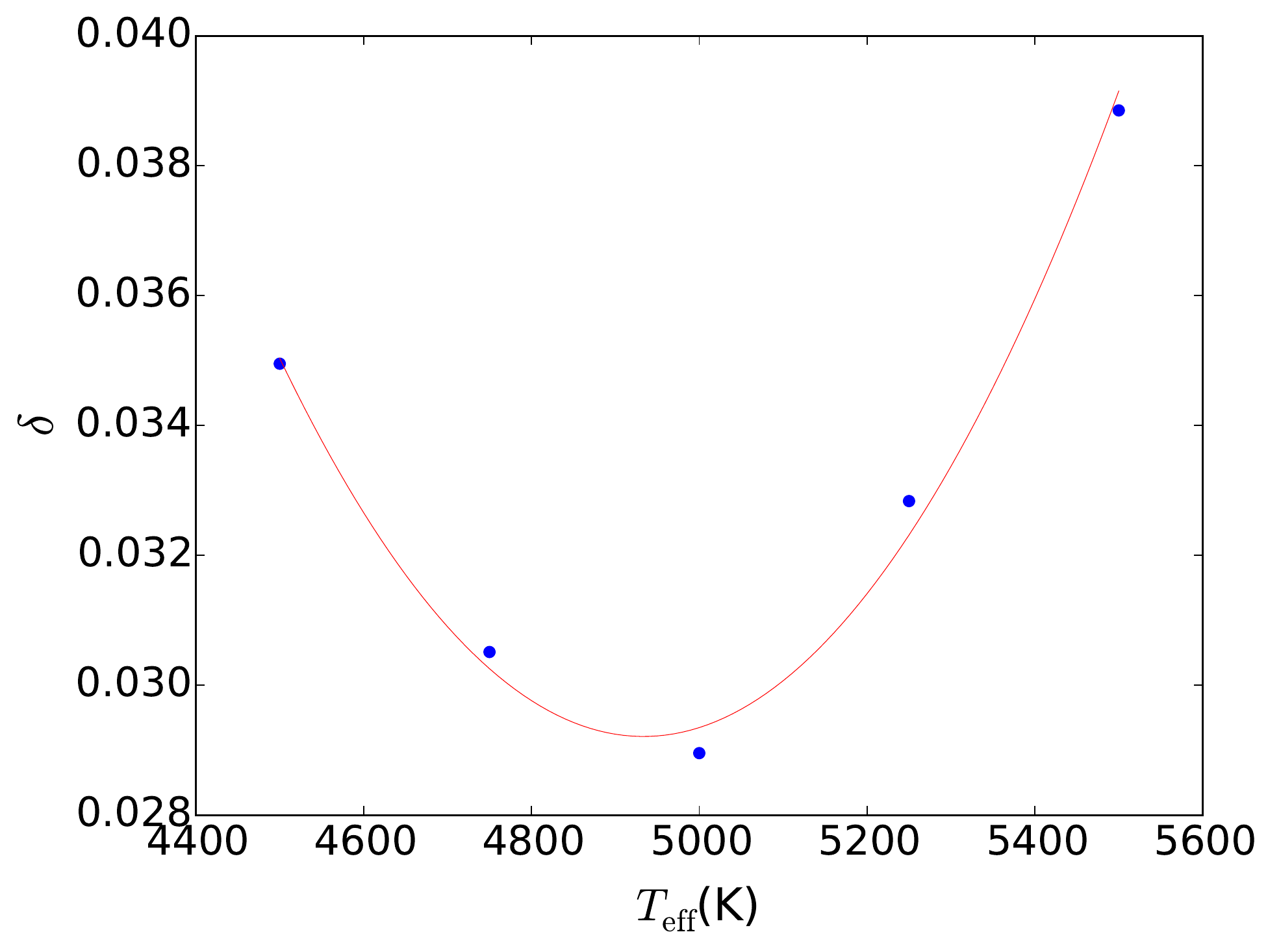}
        		}\\ %  ------- End of the first row ----------------------%
        		\subfigure
		{%
            	\label{fig:third}
            	\includegraphics[width=0.45\textwidth]{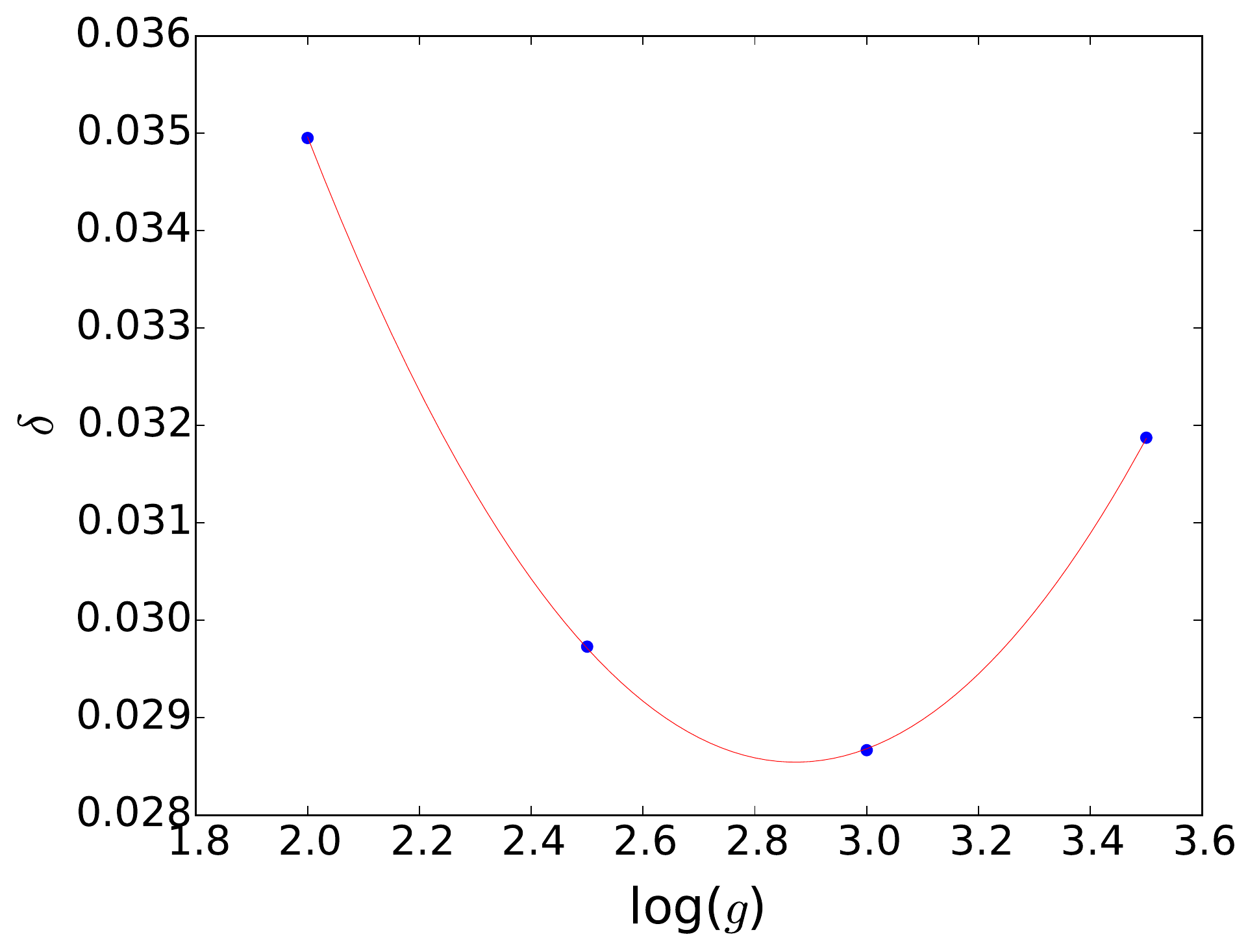}
        		}
        		\subfigure
		{%
            	\label{fig:fourth}
            	\includegraphics[width=0.45\textwidth]{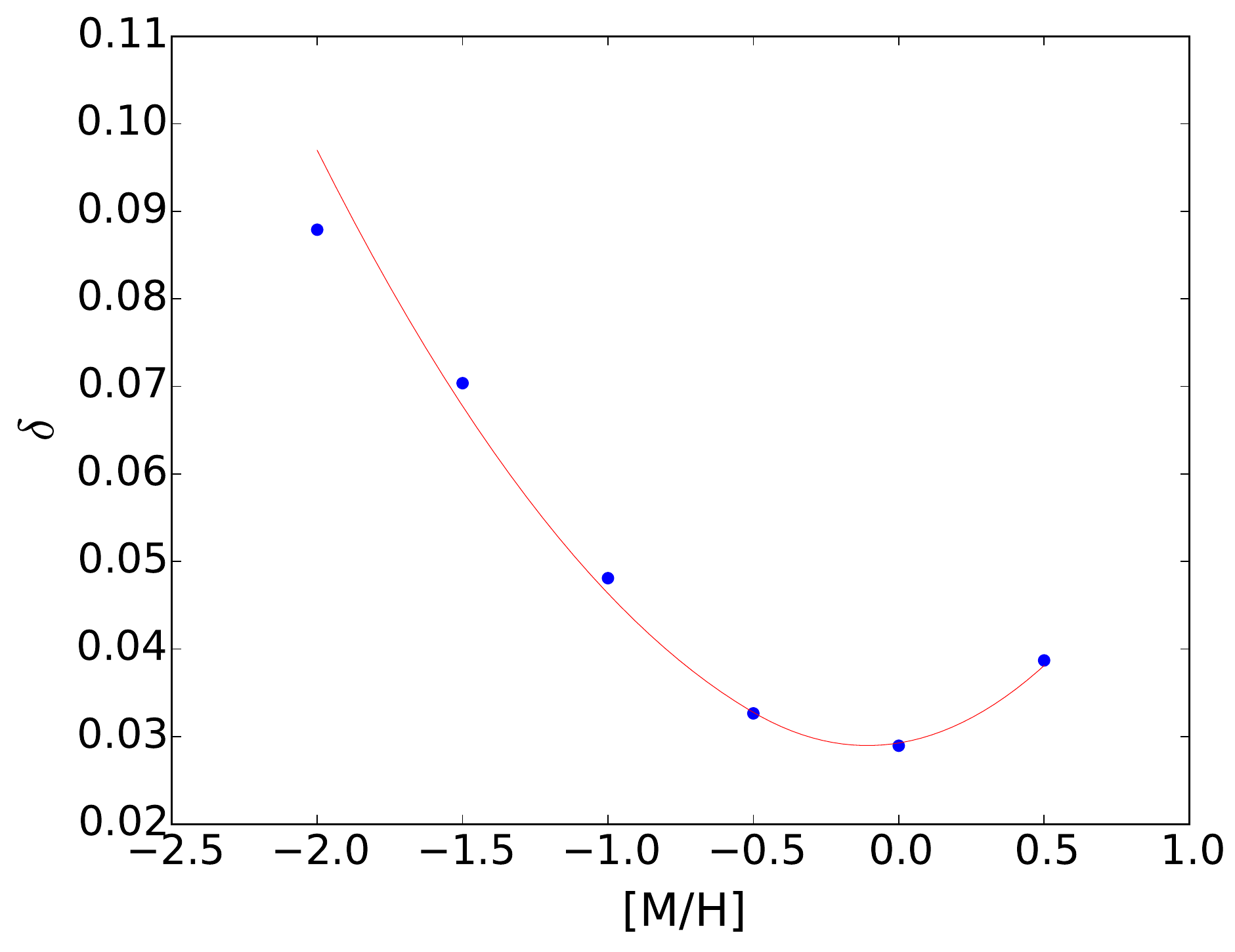}
        		}
 \end{center}
 \caption{An example of the final steps to measure stellar parameters. Upper left: six surfaces of the figure-of-merit $\delta$, each with a fixed [M/H] value; upper right: 1D B-spline fit of the minima of the five surfaces of \teff ; lower left: 1D B-spline fit of the minima of the four surfaces of \logg ; lower right: 1-D B-spline fit of the minima of the six surfaces of [M/H].}
\label{fig:Hect_spectro_step2}
\bigskip
\end{figure*}

Lastly, we interpolated $\delta$ in the 3D space with spline functions. For each grid value of a parameter, we used a 2D spline fit to find the minimum of the surface constructed by the other two parameters, then used a 1D B-spline to fit the array of minimum $\delta$ corresponding to the grid values of this parameter, assuming that the weight on each point is proportional to $1/{\delta}^2$, and then found the minimum location of the best-fit 1D curve, which gave us the best-fit value of this parameter. Figure \ref{fig:Hect_spectro_step2} illustrates this procedure. To deal with the relatively coarse grid of the synthetic library we used, we assumed that $\delta$ varies smoothly over the space among grid points. In addition, we conducted several tests in the next section to prove that our spline fitting technique produces reliable stellar parameters. Note that this method returns reliable results only when the grid point corresponding to the ``actual" stellar parameters is included in the cuboid library subset we picked for this star. So we visually examined all spline fitting results, and if the minimum of a spline fit appeared outside of the library subset, we expanded the library subset of this star until the minimum was included.

The metallicities we measure in this work are the overall metal abundances ([M/H]), according to the Kurucz synthetic library, instead of the more popular iron abundance ([Fe/H]). SPC and SME spectroscopic tools both used Kurucz model, and both output the overall metallicity [M/H] (SME also calculates the abundances of a range of individual metals), making their results easy to compare with. However, various other studies only measured the iron abundances. \citetalias{Brewer2016} addressed this problem by calculating the abundances of different metals relative to iron ($[\epsilon/\rm Fe]$), and found out that most elements' abundances are tightly correlated with the iron abundance for stars between 5000~K and 6200~K with [Fe/H]~$>-0.2$. The RMS scatter of $[\epsilon/\rm Fe]$ is between 0.12 dex and 0.25 dex, although we do see expected trends in $[\epsilon/\rm Fe]$ of $\alpha$ elements with respect to [Fe/H], where $[\epsilon/\rm Fe]$ of $\alpha$ elements increase with the increase of iron abundance. \citetalias{Dong2014} calculated [Fe/H] for \kep\ stars, and compared with the SPC [M/H] results for 47 overlapping \kep\ stars. They reported that [Fe/H] and [M/H] of these 47 stars only have a small mean difference of $-0.006\pm 0.015$ dex with no obvious trend with respect to \teff\ or \logg, and the result is reliable for stars with 4600 K$<T_{\rm eff}<$6900 K and $\rm -0.3 < [Fe/H] < 0.5$. In this work, unless otherwise noted, we assumed that all stars with which we are concerned have iron/metal ratios similar to the Sun, so that we can compare with other studies even if they only present iron abundances. 

\subsection{Spectroscopic Binary Identification}

Spectroscopic binaries are identified by cross-correlating observed spectra with their corresponding best-fit model spectra. For single stars, the cross-correlation function (CCF) should be symmetric around one peak, while for binary star systems, a secondary peak is expected in addition to the main peak in the CCF. We mirrored each cross-correlation function around its strongest peak, and took the maximum relative difference of the two sides $\rm \Delta CCF = \sup_{x}|CCF_{left}(x)-CCF_{right}(x)|/CCF_{center}$ as an indication if its level of asymmetry. If $\rm \Delta CCF> 0.1\%$, and the maximum difference corresponds to a secondary peak located within 3\AA\ from the main peak, then we consider the spectrum representing a potential binary system. Of all 776 stars, we identified 46 potential binaries. They are flagged in Table \ref{tbl-1}, but otherwise included in our analysis.

\bigskip

\section{Tests and Error estimation} \label{section4}

We started with checking the robustness of the continuum fitting. As a first approach, we took the raw Hectochelle spectrum of a Kepler star, divided it by the best-fit library spectrum corresponding to the star. If the continuum fitting was accurate, we would recover the best-fit continuum derived for the star. Therefore we refitted the resultant profile from the division with our continuum model, and found that the best-fit continuum highly coincides with the original best-fit continuum of the star derived following the procedures in Section \ref{section3}. In the second approach, we divided the raw Hectochelle spectrum with a normalized HIRES spectrum, and again found a good coincidence between the division product and our original best-fit continuum of the Hectochelle spectrum. We repeated the two tests above on several Hectochelle spectra and saw good coincidences in all cases, which is convincing evidence that our continuum fittings are robust. \\

Because $\delta$ (on the order of a few percent) on each library grid point was dominated by the uncertainty on the continuum profile, and was usually larger than the uncertainty of the observation, we could not estimate the uncertainties on stellar parameters analytically. Instead, using the same idea from \citet{Valenti2005}, we calculated empirical uncertainties on the stellar parameters by picking out stars with multiple measurements and comparing the best-fit parameters for each observation of the same stars. Using stars with multiple observations, the uncertainties on the stellar parameters were calculated within each of the following three tests, then we combined the results and assign a reasonable uncertainty value to each stellar parameter. This analysis also served as a test of the reliability of our stellar parameter estimates. 

\subsection{Reproducing Solar Properties}

As a first test, we analyzed a set of Hectochelle twilight spectra, attempting to reproduce solar parameters. The twilight spectra were observed with Hectochelle on Feb 27, 2015 with a 30 seconds exposure time, and $SNR\approx 120$. We extracted 203 available twilight spectra from the Hectochelle fibers and measure the solar parameters \teff, \logg\ and [M/H] with these 203 twilight spectra following the analysis procedure described in the last section. 

\begin{figure}
  \includegraphics[width=\linewidth]{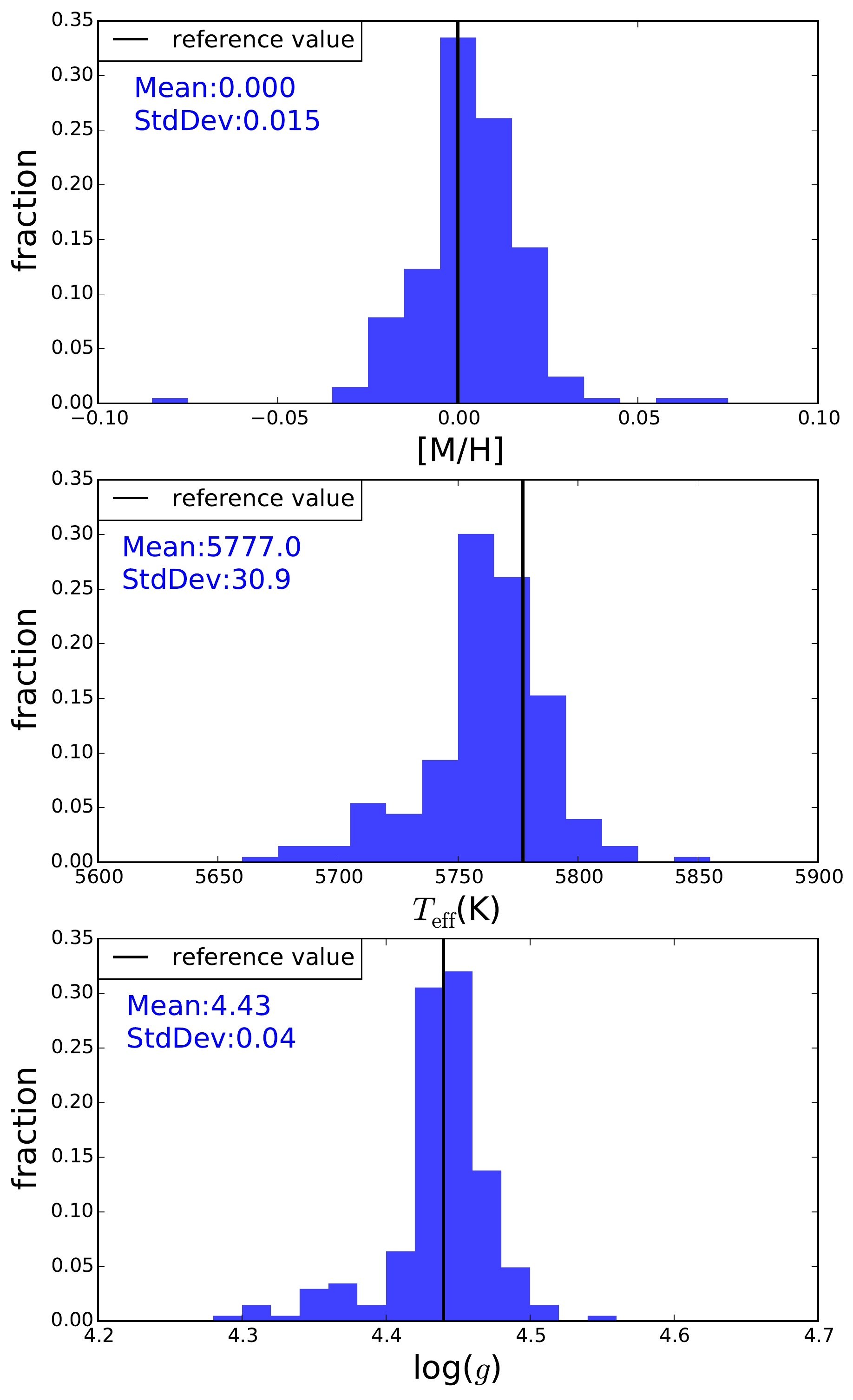}
  \caption{Measured value distributions of solar parameters obtained with 203 Hectochelle twilight spectra. The measurements are consistent with the reference values of the solar parameters given in \citetalias{Brewer2016} within $1 \sigma$ uncertainties. The reference value positions are shown with black vertical lines.}
  \label{fig:solar_prop}
  \bigskip
\end{figure}

Figure \ref{fig:solar_prop} shows the measured value distributions of the three solar parameters obtained with these 203 spectra. Based on the 203 measurements, we determined the averages and $1 \sigma$ uncertainties of \teff, \logg\ and [M/H]: $T_{\rm eff} = 5760.7\pm 26.2$ K, $\log{g} = 4.43 \pm 0.04$, [M/H] = $0.004 \pm 0.015$ dex. 

We used the reference values of solar parameters \teff\ and \logg\ from \citetalias{Brewer2016}, which are 5777~K and 4.44 respectively, and $\rm [M/H]_{\odot}$ is 0 by definition. Our method reproduces the solar properties successfully within $1 \sigma$ uncertainties. Assuming that the values given in \citetalias{Brewer2016} are the ``true'' values of \teff\ and \logg, and taking 0 as the true value of [M/H], we calculate the uncertainties on our stellar parameter measurements, which are ${\sigma}_{\Teff} = 30.9$ K, ${\sigma}_{\Logg}=0.037$, and ${\sigma}_{\MH}=0.015$ dex.

\subsection{Comparison with Other High-Resolution Spectra}

To check the reliability of our spectroscopic analysis for Hectochelle spectra, we checked for stars with Hectochelle observations that also have stellar parameter estimations from other high-resolution spectroscopy. We extracted available Hectochelle spectra of 21 HD stars, estimated their stellar parameters, and then looked up their values from previous works in the VizieR database. We found that 5 of the HD stars have reliable stellar parameter estimations: \citet{Anderson2012} reported \teff\ and [Fe/H], and \citet{Navarro2012} reported \logg\ of HD24189; \teff, \logg\ and [Fe/H] of HD10780 and HD50692 are reported by \citet{Prieto2004} and \citet{Fischer2005} respectively; and \citet{Micela1990} reported \teff, \logg\ and [Fe/H] of HD23386 and HD23352.
For all 5 HD stars only [Fe/H] are reported, however, because they are all FGK stars with metallicities between $-0.2$ and 0.2 dex, their [M/H] and [Fe/H] values should be tightly constrained around 1:1 ratio \citepalias{Brewer2016}.

\begin{figure*}
\begin{center}
		\subfigure
		{%
		\label{fig:first}
		\includegraphics[width=0.5\textwidth]{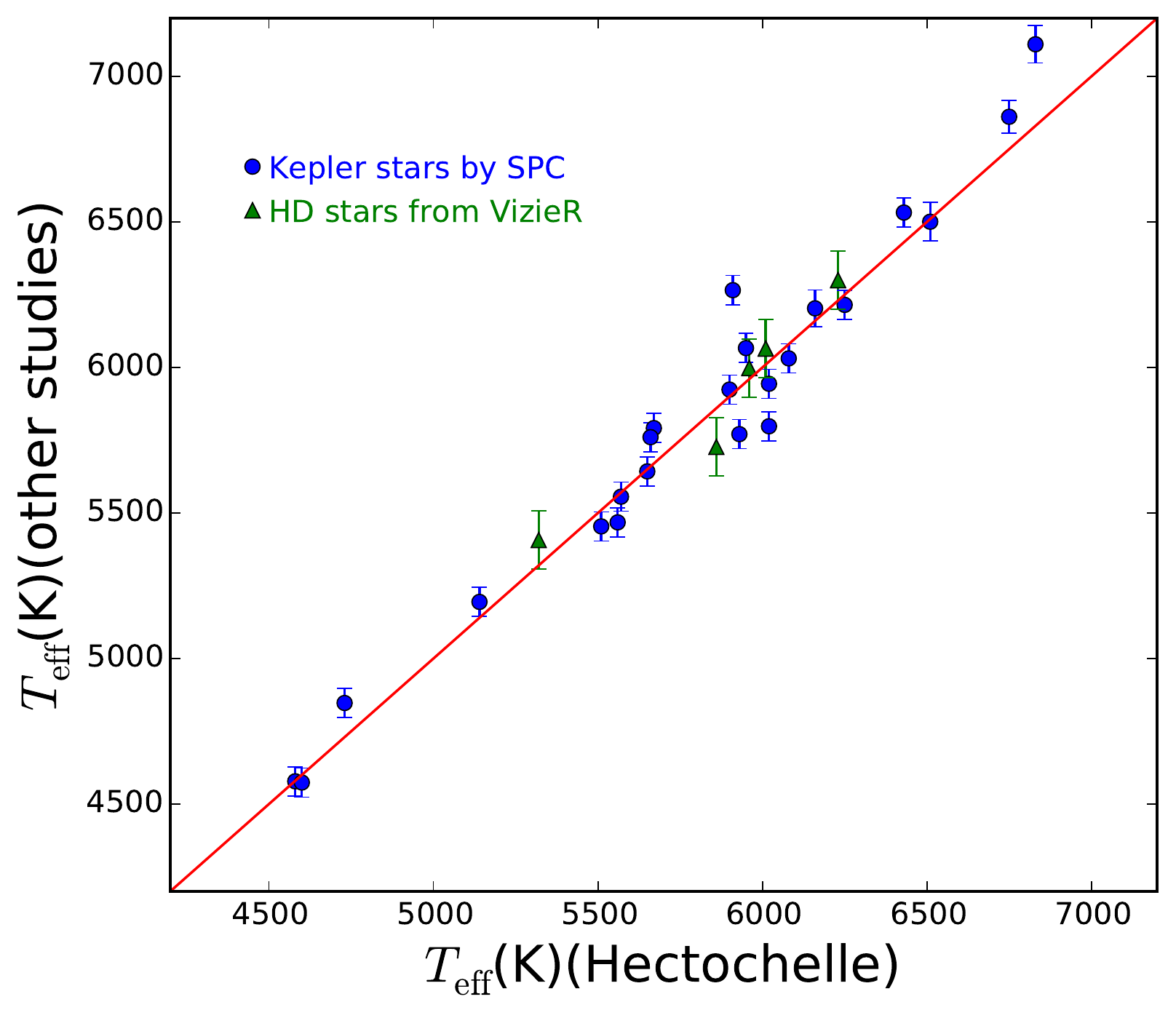}
		}%
        		\subfigure
		{%
           	\label{fig:second}
           	\includegraphics[width=0.5\textwidth]{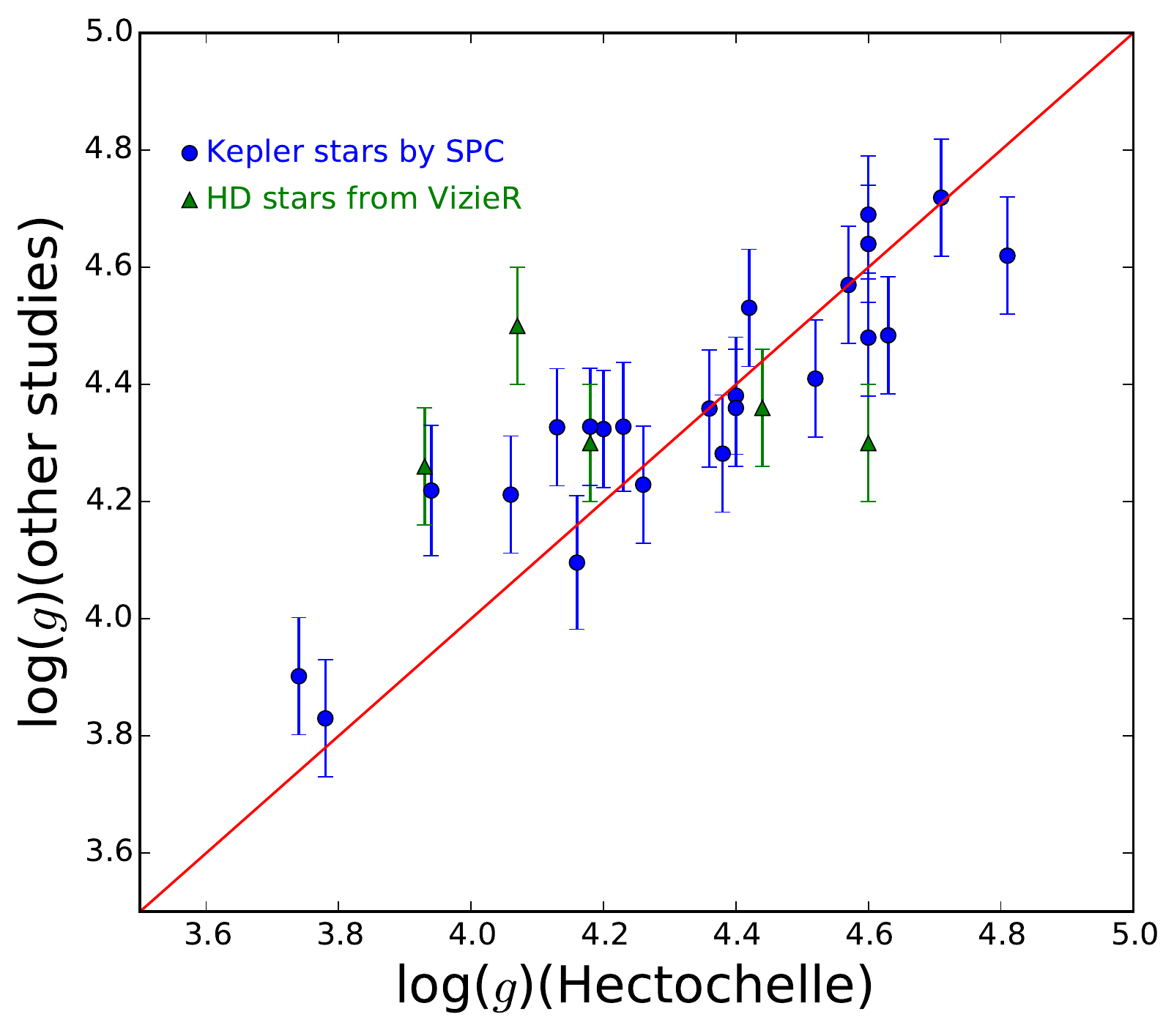}
        		} %  ------- End of the first row ----------------------%
        		\subfigure
		{%
            	\label{fig:third}
            	\includegraphics[width=0.5\textwidth]{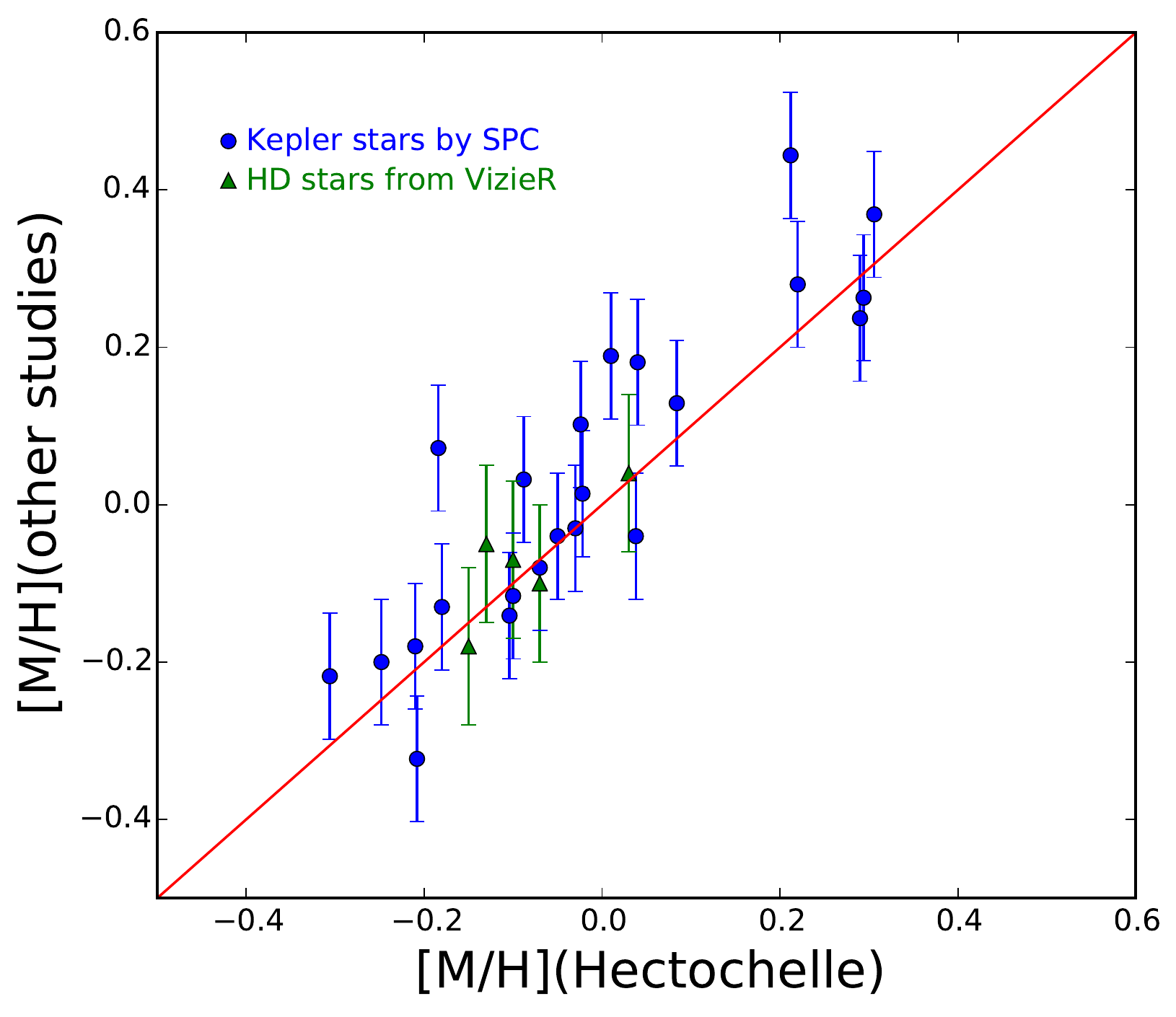}
        		}%
 	\end{center}
 	\caption{Comparison between stellar parameters measured with the Hectochelle spectra and those measured with other high-resolution data. The blue circles represent comparison between Hectochelle and SPC for 23 \kep\ stars; the green triangles represent comparison between Hectochelle and the literature measurements for 5 HD stars. \teff s are tightly constrained around the 1:1 line; \logg\ s and [M/H]s are not constrained as tightly, but agreement is reasonalbe given the uncertainties. The standard deviations of the three parameters are: $\sigma_{\Teff, {\textnormal{\textsc{spc-hecto}}}} = 122$, $\sigma_{\Logg, {\textnormal{\textsc{spc-hecto}}}} = 0.16$, and $\sigma_{\MH, {\textnormal{\textsc{spc-hecto}}}} = 0.10$.}
	\label{fig:compare_specs}
	\bigskip
\end{figure*}

We picked out the 6 \kep\ stars in our sample that were also analyzed with SPC, and compare our results. To broaden the comparison sample, we observed 17 more \kep\ stars from our sample with the Tillinghast Reflector Echelle Spectrograph (TRES) and analyzed them with the SPC tool. TRES has a resolution of R$\approx$44000, covering a wavelength range from 3800~\AA\ to 9100~\AA. The 17 \kep\ stars were observed with the exposure times between 300 seconds and 3600 seconds, and their SNR are between 25 and 50. Therefore the final sample that was compared with the SPC results consists of 23 \kep\ stars spanning a \teff\ range from 4200~K to 7000~K, a \logg\ range from 3.5 to 5.0 and a [M/H] range from $-0.5$ dex to 0.5 dex.

Figure \ref{fig:compare_specs} presents the comparison between our Hectochelle parameters and those from the SPC or the literature. The effective temperatures are tightly constrained around the 1:1 line across the \teff\ range from 4500~K to 7000~K, with a standard deviation of $\sigma_{\Teff, {\textnormal{\textsc{spc-hecto}}}} = 122$~K; \logg\ values show a larger scatter, indicating the known lack of precision in spectroscopic \logg\ measurements. Nonetheless, all our \logg\ data points are distributed in the vicinity of the 1:1 ratio line with a deviation of $\sigma_{\Logg, {\textnormal{\textsc{spc-hecto}}}} = 0.16$. [M/H]s are relatively well constrained with a standard deviation of $\sigma_{\MH, {\textnormal{\textsc{spc-hecto}}}} = 0.10$ dex, indicating that our [M/H] measurements are reliable. Combining with the uncertainties on SPC measurements of stellar parameters, which are $\sigma_{\Teff, {\textnormal{\textsc{spc}}}} = 50$ K, $\sigma_{\Logg, {\textnormal{\textsc{spc}}}} = 0.1$ and $\sigma_{\MH, {\textnormal{\textsc{spc}}}} = 0.08$ dex, and assuming that the uncertainties on the spectroscopic measurements of the 5 HD stars obtained from VizieR are approximately the same as SPC, we calculate the uncertainties on our stellar measurements with the equation $\sigma_{\textnormal{\textsc{hecto}}}=\sqrt{{\sigma_{\textnormal{\textsc{spc-hecto}}}}^2+{\sigma_{\textnormal{\textsc{spc}}}}^2}$. The results are $\sigma_{\Teff}= 132$ K, $\sigma_{\Logg}= 0.19$ and $\sigma_{\MH}= 0.13$ dex.

\subsection{NGC 752}

Stars in a cluster share approximately the same age and metallicity. Therefore by measuring metallicities for multiple stars in a cluster and calculating their standard deviation, we can estimate the uncertainty on our metallicity measurements. NGC 752 is an open cluster approximately 460 pc from us in the constellation Andromeda with an age of about 1.45 Gyr, and it consists of stars with $V$ magnitude 8.6 or fainter \citep{Twarog2015}. It was observed with Hectochelle to determine membership and multiplicity in support of a program seeking to study the chromospheric activity of its solar-type stars. Our detailed analysis of the cluster membership and its properties will be presented in a separate work, but for now, we can validate our spectroscopic analysis procedure by applying it to the available Hectochelle spectra of NGC 752 members and testing for systematic trends and metallicity dispersions.  
The full sample consists of 130 stars; for this analysis, we selected 36 stars that appear to be single members based on the radial velocities and photometric proximity to the cluster isochrone solution in a color--magnitude diagram. The selected stars have $V$ magnitudes between 9.0 and 14.8, and temperature between 4500~K and 6500~K. Observations were taken using an exposure time of 1200 seconds, resulting in a typical $SNR$ of $\approx$70. We perform our analysis on these stars as above to obtain their \teff, \logg , and [M/H]. In addition, we obtain photometric measurements on \teff\ for 17 stars in our sample by isochrone fitting, and compare with our spectroscopic measurements to test the reliability of our measurements. 

\citet{Hobbs1992} determined the metallicity of NGC 752 to be [Fe/H] = $-0.09\pm 0.05$ by measuring 8 main-sequence stars; \citet{Sestito2004} found a metallicity of [Fe/H] = $0.01\pm 0.04$ from a sample of 18 stars; \citet{Carrera2011} derived a metallicity of [Fe/H] = $0.08\pm 0.04$; and \citet{Reddy2012} compute an [Fe/H] of $-0.02\pm 0.05$ from 4 giant-star members. Although all these previous studies focused on the iron abundance of NGC 752, it is reasonable to compare our [M/H] results with their iron abundances because NGC 752 has a slightly sub-solar metallicity (\textgreater -0.2), and we only compare FGK type stars. Thus we calculated the average and standard deviation of the previous measurements on the metallicity of NGC 752 and found that $\rm [M/H]_{NGC 752} = -0.005 \pm 0.045$.

\begin{figure} 
  \includegraphics[width=\linewidth]{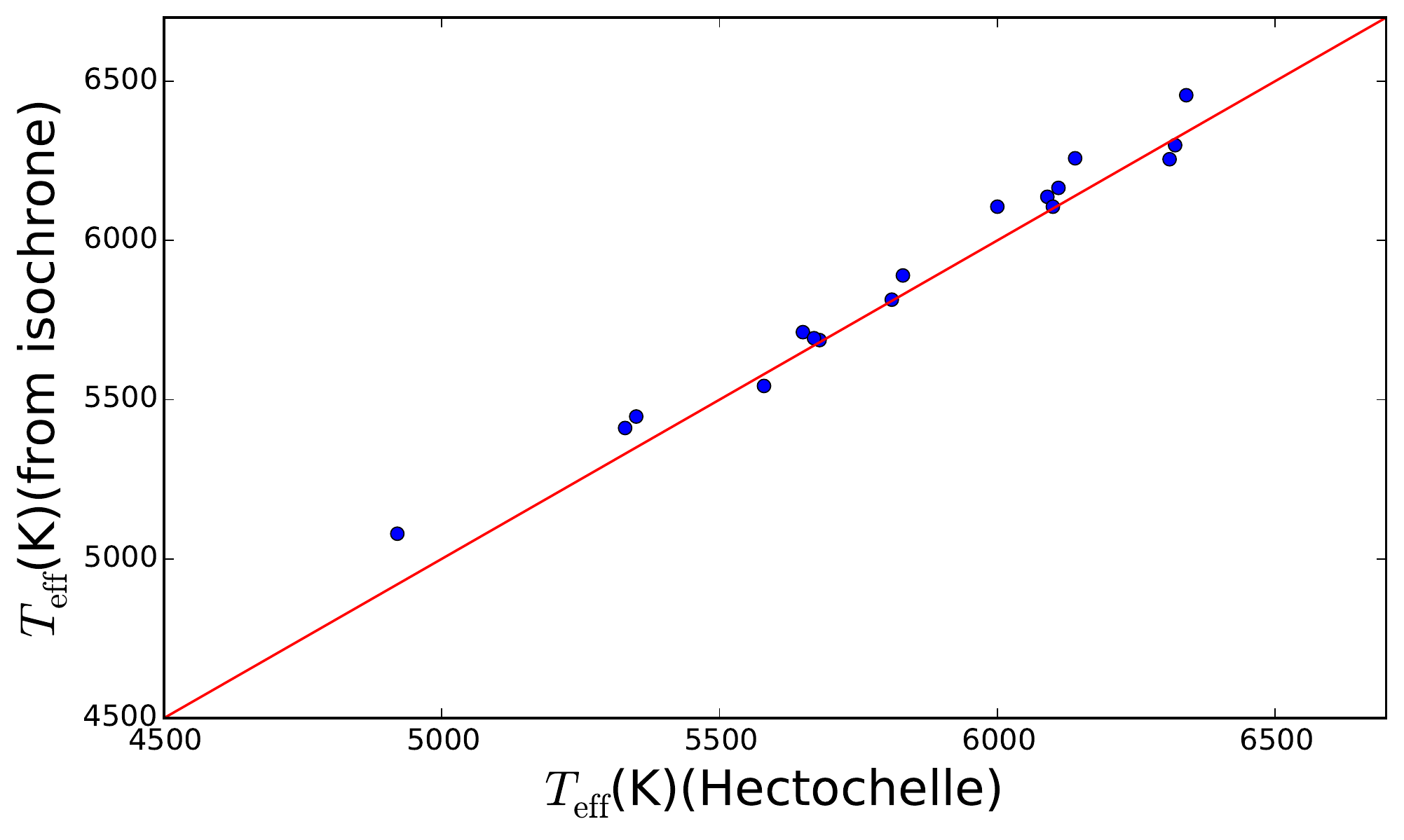}
  \caption{Comparison of \teff\ determined with Hectochelle spectra and \teff\ determined with photometry using the isochrone for a 1.5-Gyr-old cluster with $\rm [Fe/H] = 0.0$. All the data are well constrained to the 1:1 line with $\sigma_{\Teff}<100$K. Although the photometric \teff\ is systematic higher than Hectochelle spectroscopic \teff, this effect is relatively small.}
  \label{fig:NGC752_Teff}
  \bigskip
\end{figure}

Figure \ref{fig:NGC752_Teff} shows the comparison between the \teff s of 17 NGC 752 stars measured with Hectochelle spectroscopy and those measured with photometry using the isochrone of a 1.5-Gyr-old cluster with [Fe/H] = 0.0. 

The effective temperatures are consistent in general on the range from 4500~K to 6500~K, with a standard deviation smaller than 100~K, although photometric \teff\ are generally slightly higher than the Hectochelle spectroscopic \teff. Previous studies on NGC 752 have suggested a range of possible ages from 1.2 Gyr to 1.9 Gyr \citep{Bilir2006, Bartasiute2007} and a slightly sub-solar metallicity, so by assuming a 1.5-Gyr-old solar metallicity isochrone gives a systematic offset towards a younger and slightly more metal rich cluster. This in turn could result in overestimated photometric \teff s. In addition, spectroscopic measurements on \teff\ are known to be slightly lower than the photometric measurements, which we discuss in Section \ref{section5}. 

\begin{figure}
  \includegraphics[width=\linewidth]{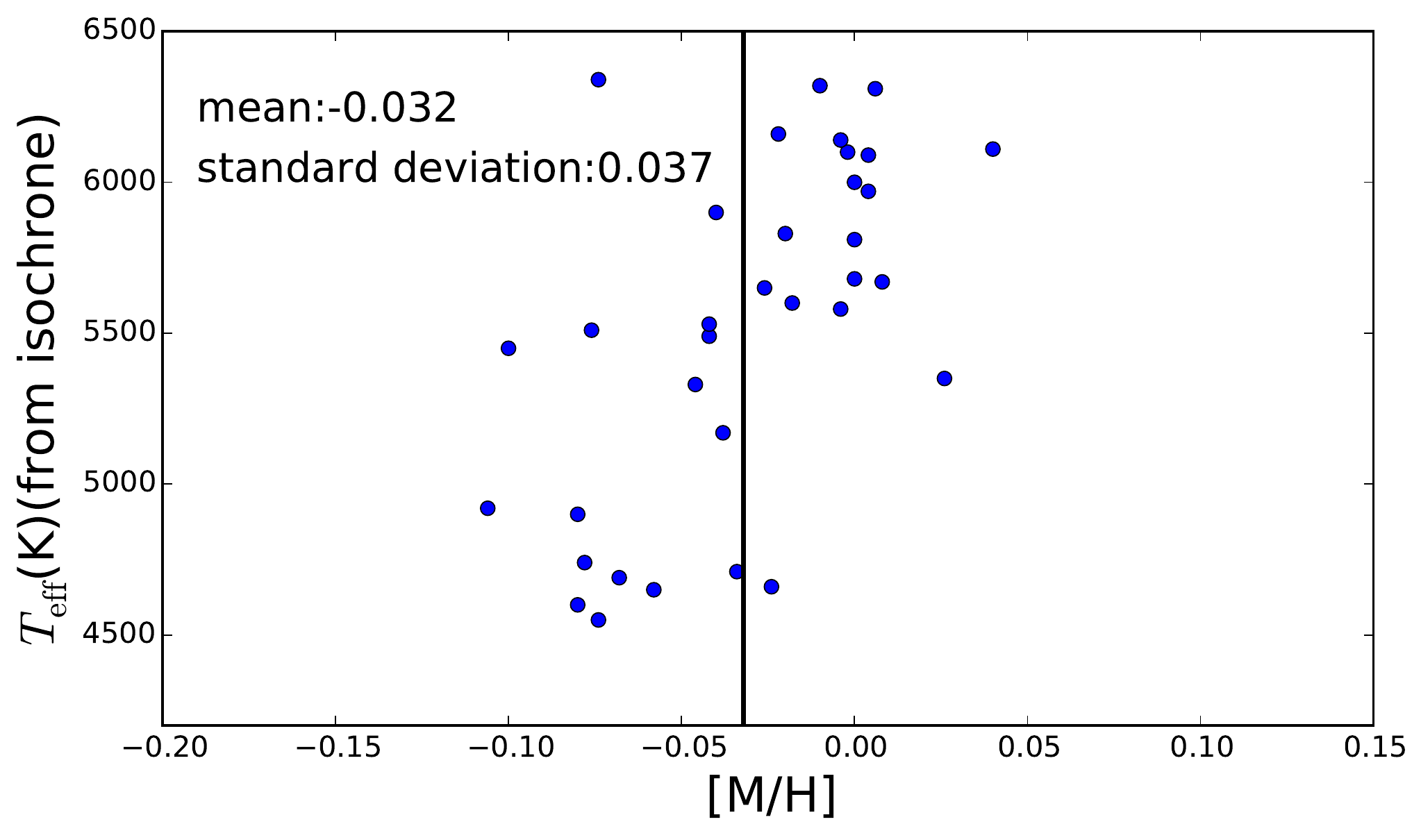}
  \caption{Metallicities of stars in the NGC 752 sample, determined with Hectochelle spectroscopy. Mean and standard deviation of the distribution are -0.032 dex and 0.037 dex respectively, which is consistent with most of previous studies on NGC 752.}
  \label{fig:NGC752_Z}
  \bigskip
\end{figure}

Figure \ref{fig:NGC752_Z} presents the [M/H] distribution of our 36 NGC 752 stars, which shows [M/H] = $-0.032 \pm 0.037$. The slight decreasing trend in [M/H] for stars cooler than 5500~K is also observed in \citet{Brewer2016}, which is common for spectroscopic analyses using Kurucz model. Since this model-induced trend is not significant ([M/H] of a 5000~K star is only shifted lower by around 0.1 dex, according to \citet{Brewer2016}), and most (about 65\%) stars in our Kepler sample are hotter than 5500~K, the influence on the final metallicity distribution of the Kepler sample is negligible. In addition, the mean [M/H] of our NGC 752 sample is consistent with the results from all the previous works except for \citet{Carrera2011}, which may suffer from small number statistics. Based on the test on NGC 752, we estimate that the uncertainty on our [M/H] measurement is $\sigma_{\MH} = 0.037$.

\subsection{Uncertainty Assignment}

All our tests give relatively consistent error estimates. The test on 203 solar spectra suggests errors on our stellar parameters are ${\sigma}_{\Teff} = 30.9$ K, ${\sigma}_{\Logg}=0.037$, and ${\sigma}_{\MH}=0.015$ dex; the comparison with other high-resolution spectroscopy indicates uncertainties of $\sigma_{\Teff}= 132$ K, $\sigma_{\Logg}= 0.19$, and $\sigma_{\MH}= 0.13$ dex; and the test on NGC 752 favors metallicity uncertainties of $\sigma_{\MH} = 0.037$. In light of these 3 tests above, we conclude that the uncertainties of our \teff, \logg\ and [M/H] measurements are well represented with $\sigma_{\Teff}=100$ K, $\sigma_{\Logg}=0.1$ and $\sigma_{\MH}=0.1$, which we adopt as our formal errors.

\bigskip

\section{Results} \label{section5} 

We report the spectroscopic parameters of the 776 \kep\ stars in our sample in Table \ref{tbl-1}. Our measurements of effective temperatures agree well with those presented in the KSPA for stars with $T_{\rm eff}\lesssim 5500$ K, while for stars with $T_{\rm eff}\gtrsim 5500$ K, the effective temperatures given by the KSPA are systematically higher than our spectroscopic temperatures. The systematic difference is around 200~K at 6000~K, and hotter stars have slightly larger systematic differences, while cooler stars have smaller differences. This deviation is expected since $\approx$70\% of the stellar parameter values in the KSPA are from the KIC photometric estimates, and photometrically derived \teff\ are known to be systematically higher than those determined from spectroscopy \citep{Frebel2013, Hollek2011}. The surface gravities of the 776 \kep\ stars are much better constrained by our spectroscopic analysis than by the KIC photometric analysis.

\subsection{Metallicity Distribution of the \kep\ Sample} \label{section5_1}

\begin{figure}
  \includegraphics[width=\linewidth]{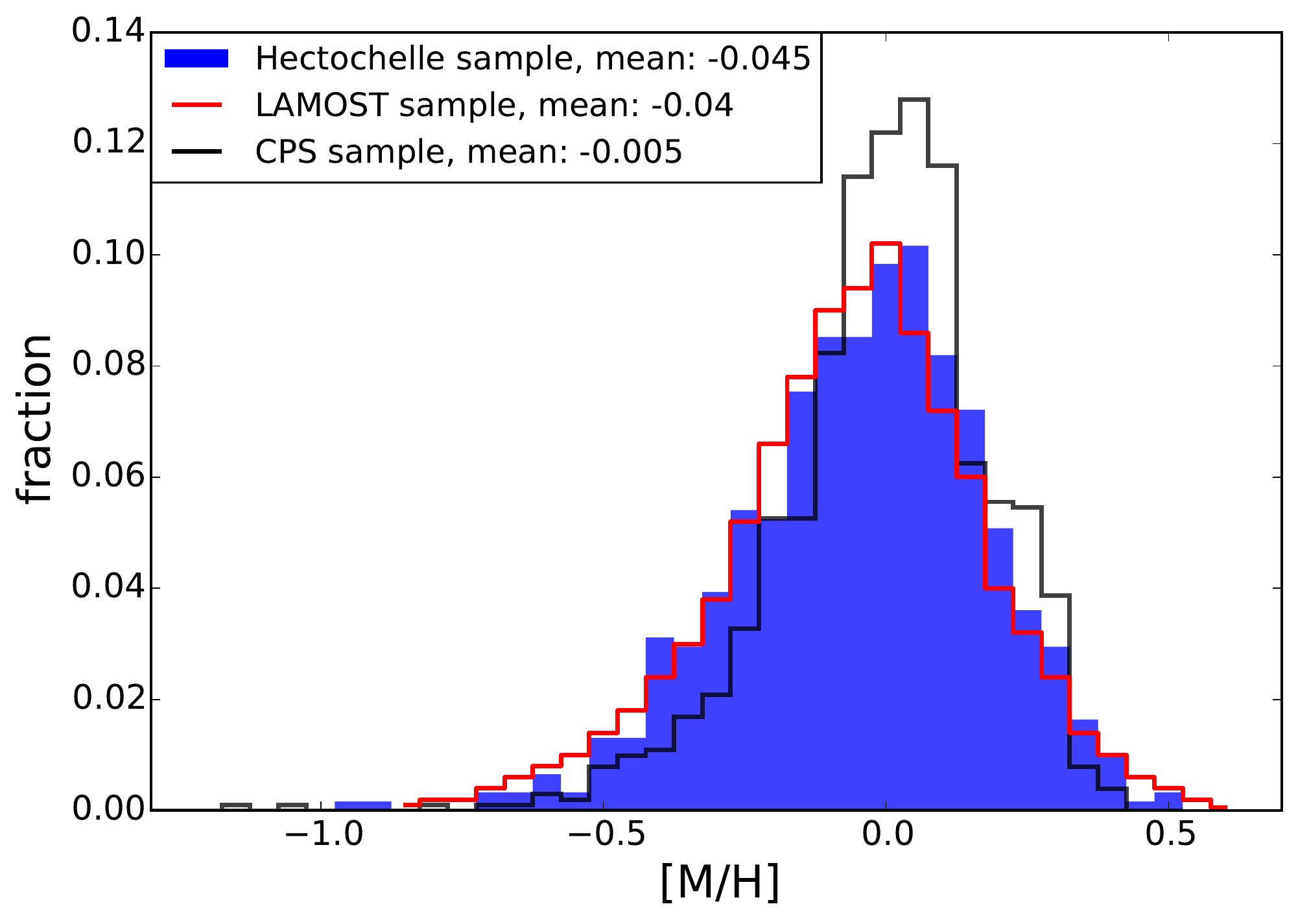}
  \caption{The [M/H] distribution of all 610 dwarf stars (with \logg\ $>3.5$) in our \kep\ sample is shown with the blue histogram. The red histogram edge shows the [Fe/H] distribution determined with the LAMOST low-resolution sample of 14000 \kep\ dwarf stars, and the black histogram edge shows the [M/H] distribution of 1008 CPS dwarf stars derived with SME.}
  \label{fig:stat_Z}
  \bigskip
\end{figure}

We report the metallicity distribution of dwarf stars in the \kep\ field in Figure \ref{fig:stat_Z}, which is composed from 610 dwarf stars in our sample defined with \logg\ $>3.5$. 
The metallicity of the dwarf star sample has a mean of $\rm [M/H]_{dwarf}=-0.045\pm 0.009$ with a standard deviation of 0.225 dex. Since the Kurucz library grid has an upper limit of metallicity at $\rm [M/H]=0.5$, there are several stars piling up at $\rm [M/H]=0.5$, giving rise to the small peak on the right edge of the distribution. This does not affect the statistical result because only a small fraction ($\lesssim 1\%$) of the sample stars have metallicities $\geq 0.5$ dex. 
If we also count in the subgiant/giant stars, the metallicity distribution of our whole sample of 776 stars has a mean of $\rm [M/H]_{all}=-0.053\pm 0.008$ with a standard deviation of 0.228 dex.

\begin{figure}
  \includegraphics[width=\linewidth]{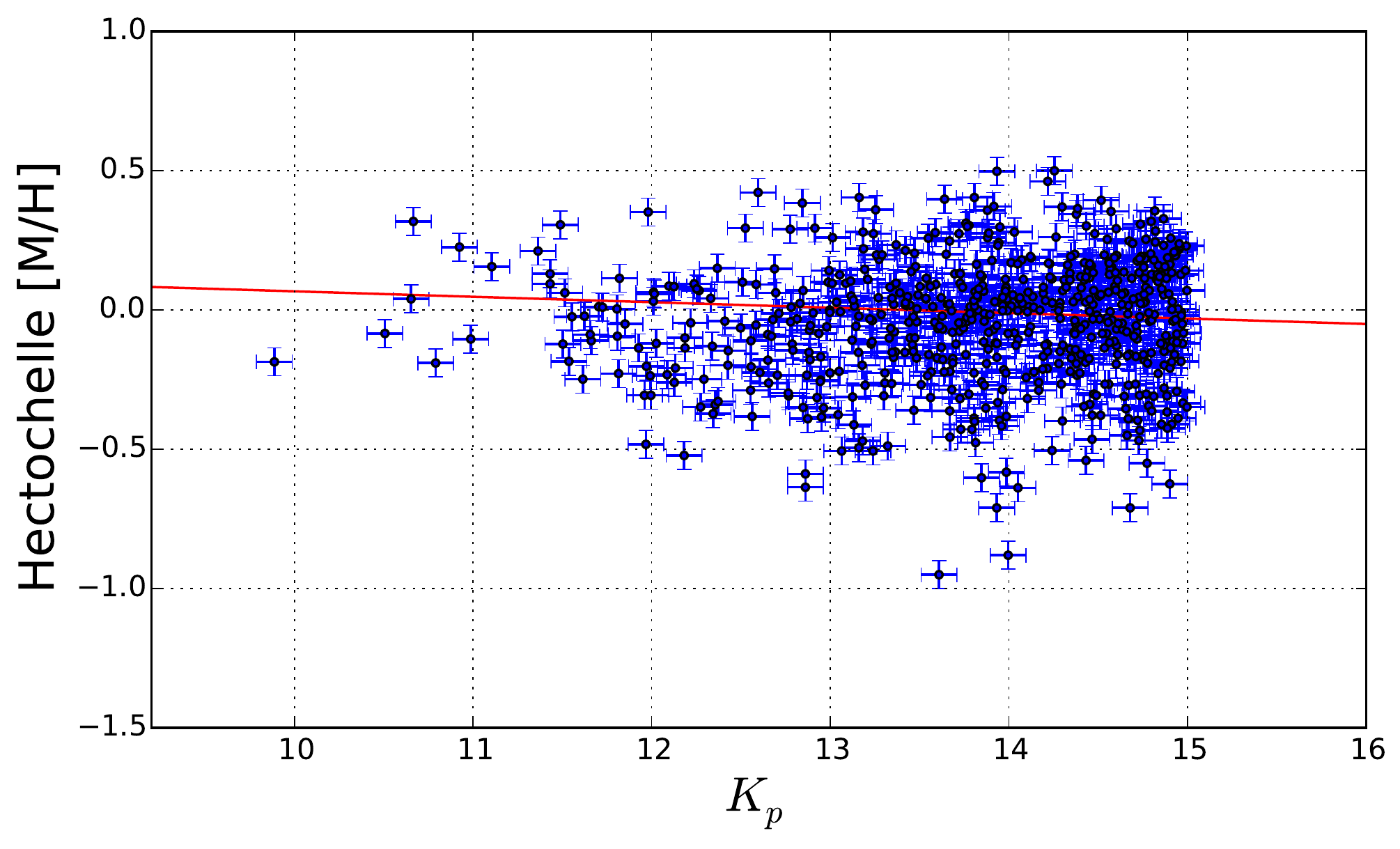}
  \caption{The distribution of [M/H] as a function of $K_p$ for stars in our sample. The red line represents the linear fit result performed with stars in our sample with $14<K_p<15$, which is the magnitude region designed for statistical studies. The best-fit slope is $-0.02\pm 0.15$, and the extrapolated [M/H] at $K_p =16$ is $-0.05\pm 0.21$ dex. Therefore the magnitude cut of $K_p<15$ of our sample can only bias the [M/H] insignificantly, with the mean [M/H] shifted by less than 0.005 dex.}
  \label{fig:Z_Kp}
  \bigskip
\end{figure}

Our \kep\ sample is magnitude limited, where all stars have $K_p<15$. Therefore there might be a systematic bias towards slightly higher metallicity since brighter stars are likely to be in the thin disc and hence more metal-rich. To investigate this effect, we fit the stellar metallicities as a function of their $K_p$, as is shown in Figure \ref{fig:Z_Kp}. The stellar population within $14<K_p<15$ were selected for statistical studies while brighter stars were selected for follow-ups, thus we perform the fit only on dwarf stars with $14<K_p<15$, although the result is unchanged if we use the full range of $K_p$. The best-fit relation is [M/H]$~=-0.02 (K_p - 16) -0.05$. The data points on Figure \ref{fig:Z_Kp} show no clear decline of [M/H] with the increase of $K_p$, and the best-fit slope of $-0.02\pm 0.15$ implies that the relation between [M/H] and $K_p$ is insignificant. As an extrapolation of the fit, the metallicity at $K_p=16$ is [M/H]$~=-0.05\pm 0.21$. And since the mean of our sample of \kep\ dwarf stars is $\rm [M/H]_{dwarf}=-0.045\pm 0.009$, the systematic shift of the [M/H] distribution because of the magnitude limit should be less than 0.005 dex. 

\subsection{Comparison with Previous \kep\ Metallicity Studies}

\begin{figure}
  \includegraphics[width=\linewidth]{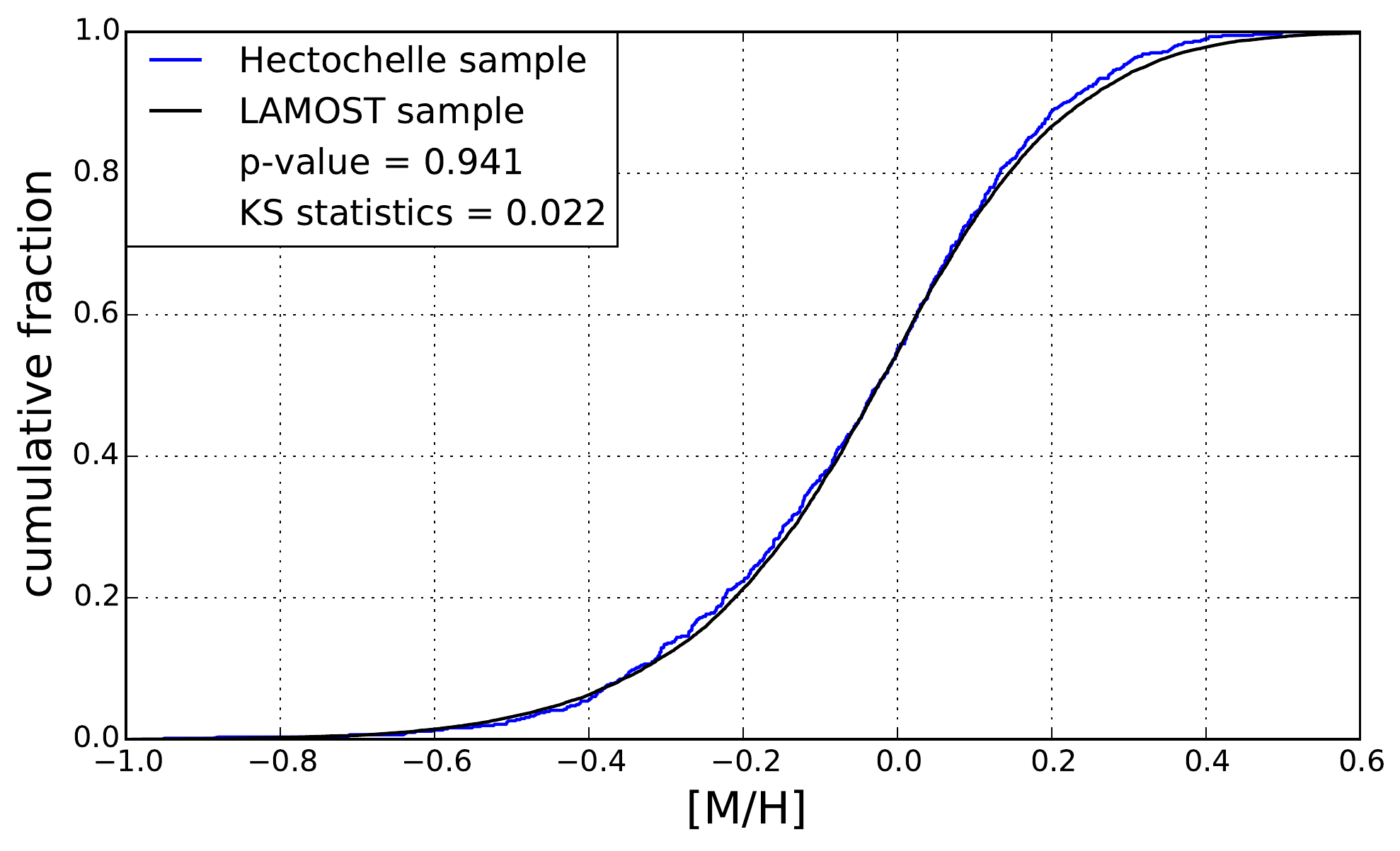}
  \caption{The two-sample KS test between the [M/H] distribution of our \kep\ sample of 610 dwarf stars observed with Hectochelle high-resolution spectroscopy and the [Fe/H] distribution of \citet{Dong2014} \kep\ sample of 14000 dwarf stars observed with LAMOST low-resolution spectroscopy. A p-value of 0.941 indicates that the two metallicity distributions of the \kep\ field are consistent.}
  \label{fig:KS_LAMOST}
  \bigskip
\end{figure}

\citetalias{Dong2014} reported the \kep\ field iron abundance distribution by measuring [Fe/H] values of 14000 \kep\ stars using the LAMOST spectroscopic survey. The [Fe/H] distribution of the LAMOST sample has a mean value of $\rm [Fe/H]=-0.040\pm 0.002$ and a 0.25 dex standard deviation, within 1$\sigma$ of our own determination. We also perform a two-sample Kolmogorov--Smirnov (KS) test, which is shown in Figure \ref{fig:KS_LAMOST}. The test has a p-value of 0.94, suggesting our [Fe/H] determinations and resulting conclusions about the overall metallicity distribution are consistent with those from \citetalias{Dong2014}. 

Although the LAMOST analysis measures iron abundance instead of overall metallicity, \citetalias{Dong2014} compared their [Fe/H] measurements of 47 stars with the SPC [M/H] measurements of the same stars, and found a mean difference of $-0.006 \pm 0.015$ dex on the [M/H] range from -0.5 to 0.5 dex, which covers most of the LAMOST sample stars. Thus we argue that the comparison between our [M/H] and the LAMOST [Fe/H] distributions is valid. 

\citet{Santerne2016} compiled 37 \kep\ dwarf stars observed with the SOPHIE high-resolution velocimetry and presented their spectroscopic parameters measured with the Equivalent Width method using either the MOOG \citep{Sneden1973} or the VWA \citep{Bruntt2002, Bruntt2004} software. They found that the average [Fe/H] value of \kep\ stars measured with the SOPHIE high-resolution velocimetry is $0.17\pm 0.04$ dex higher than that presented in \citetalias{Huber2014q16}. Because the median \kep\ stellar metallicity estimated in \citetalias{Huber2014q16} is approximately $-0.18$ \citep{Santerne2016}, we conclude that the mean metallicity of \kep\ dwarf stars is indeed only slightly sub-solar, consistent with our result. 

\subsection{Comparison with the Solar Neighborhood Metallicity Distribution}

We compare the metallicity distribution of the \kep\ star sample to that of the solar neighborhood stars. The metallicity distribution of solar neighborhood stars was obtained by combining metallicity results of 1040 CPS FGK stars presented in \citet{Valenti2005} using the SME package. Updated metallicity measurements of 1626 CPS stars are presented in \citetalias{Brewer2016} using a modified version SME. We take only dwarf stars with \logg $>3.5$, visual magnitude $V<8$ and color index $\rm B-V<1.2$, to be consistent with \citetalias{Wright2012}, then exclude any stars from the N2K program which specifically targeted metal-rich stars \citep{Robinson2007}. From this we obtain a sample of 1008 CPS stars, with a mean of $\rm [M/H]_{CPS} = -0.005 \pm 0.006$ and $\sigma_{\rm{[M/H]}}=0.187$ dex standard deviation. This distribution represents the most accurate and precise measurements of the CPS sample metallicity, and is shown in Figure \ref{fig:stat_Z}.

\begin{figure}
  \includegraphics[width=\linewidth]{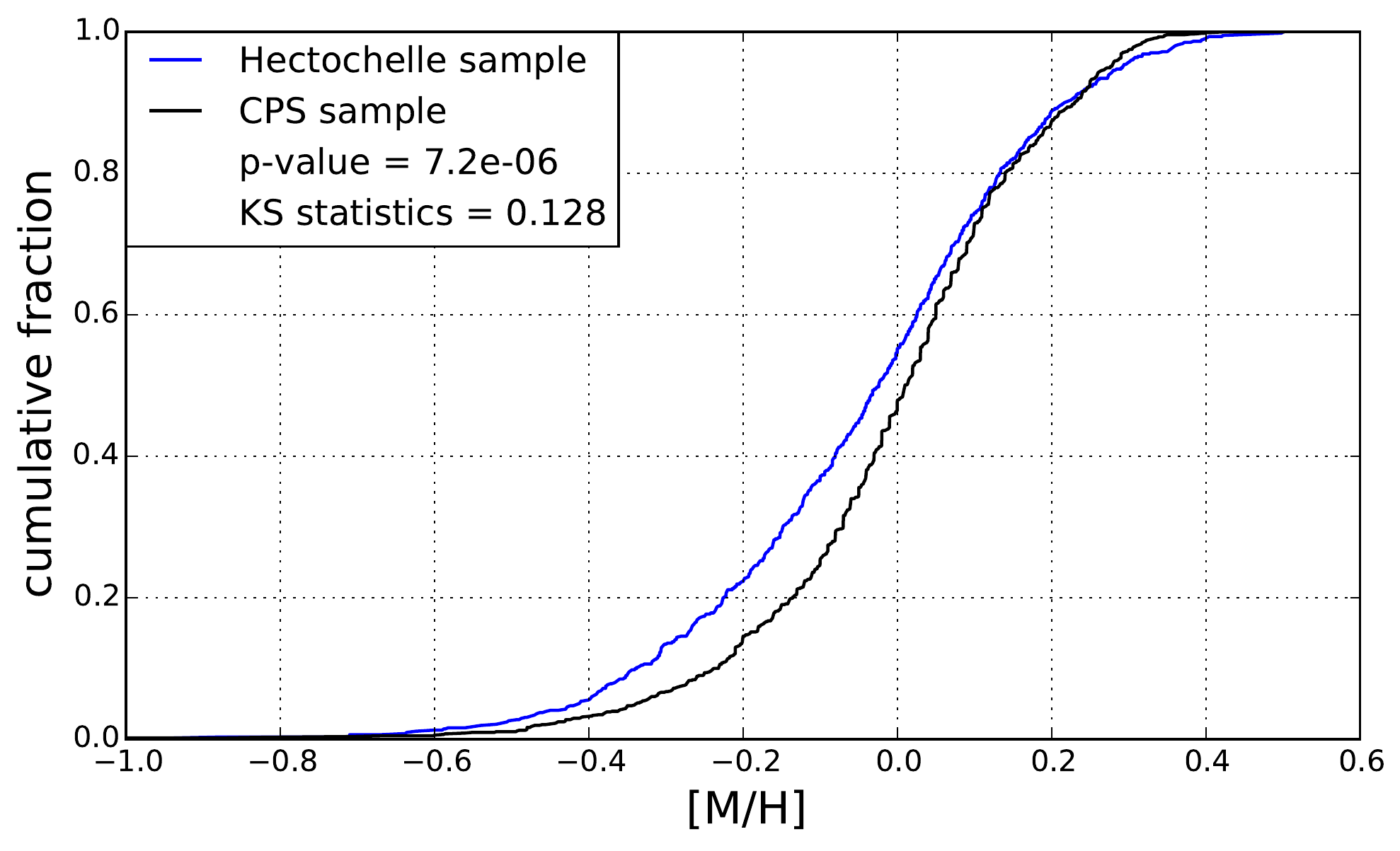}
  \caption{The two-sample KS test between the [M/H] distribution of our \kep\ sample of 610 dwarf stars and that of the CPS sample of 1008 dwarf stars. A p-value of $7.2\times 10^{-6}$ indicates that the sample of \kep\ field stars and the sample of solar-neighborhood stars are distinct. And the difference of the metallicity distributions is at least partly responsible for the HJ rate discrepancy.}
  \label{fig:KS_CPS}
  \bigskip
\end{figure}

There is a higher fraction of metal-rich stars and a slight deficiency of metal-poor stars in the CPS metallicity distribution (see Figure \ref{fig:stat_Z}). As before, we performed a two-sample KS test to compare the two [M/H] distributions quantitatively, which we report in Figure \ref{fig:KS_CPS}. A p-value of $7.2\times 10^{-6}$ rejects the null hypothesis that the two metallicity samples are drawn from the same distribution beyond 3$\sigma$ level, indicating that the sample of solar-neighbourhood stars and the sample of the \kep\ stars are distinct. 

However, the difference between the two distributions is relatively small: the CPS mean [M/H] is only $0.040\pm 0.015$ dex higher than the \kep\ mean [M/H]. While the CPS sample has an excess number of solar metallicity stars and fewer sub-solar metallicity stars, the \kep\ sample slightly outnumbers the CPS sample in stars with [M/H] between 0.35 and 0.5. Moreover, \citet{Sousa2008} found the mean iron abundance of 451 nearby stars to be $-0.09$ dex with the HARPS GTO planet search program.
Therefore it is unclear whether the difference between the metallicity distributions of \kep\ dwarf stars and solar-neighborhood dwarf stars is able to totally account for the HJ rate discrepancy. 

\subsection{HJ Occurrence Rates from the Metallicity Distributions} \label{section5_3}

With the established metallicity distributions of \kep\ stars and CPS stars, we calculate the expected HJ rates of the two surveys, using exponential relations between HJ occurrence probability and the host star's metallicity according to previous studies \citep[][J10]{Fischer2005}.

\citet{Fischer2005} reported the relation between the probability for an FGK dwarf star to host a close-in giant planet and the host star metallicity to be $P(\rm planet)=0.03\times10^{2.0\rm [Fe/H]}$. With increased Doppler precision, \citetalias{Johnson2010} confirmed the exponential relation and re-derived the exponential index, showing that: $P(\rm HJ) \propto 10^{1.2\rm [Fe/H]}$. In this section, we refit the exponential relation in two cases using our combined sample of 1008 CPS stars: 1) with the exponential index restrained to the \citetalias{Johnson2010} result, 2) with the exponential index relaxed. For each best-fit relation, we calculate the expected HJ rates of the \kep\ sample and the CPS sample.

\subsubsection{Exponential Relation with Restrained Index}

\begin{figure*}
\begin{center}
		\subfigure
		{%
		\label{fig:first}
		\includegraphics[width=0.5\textwidth]{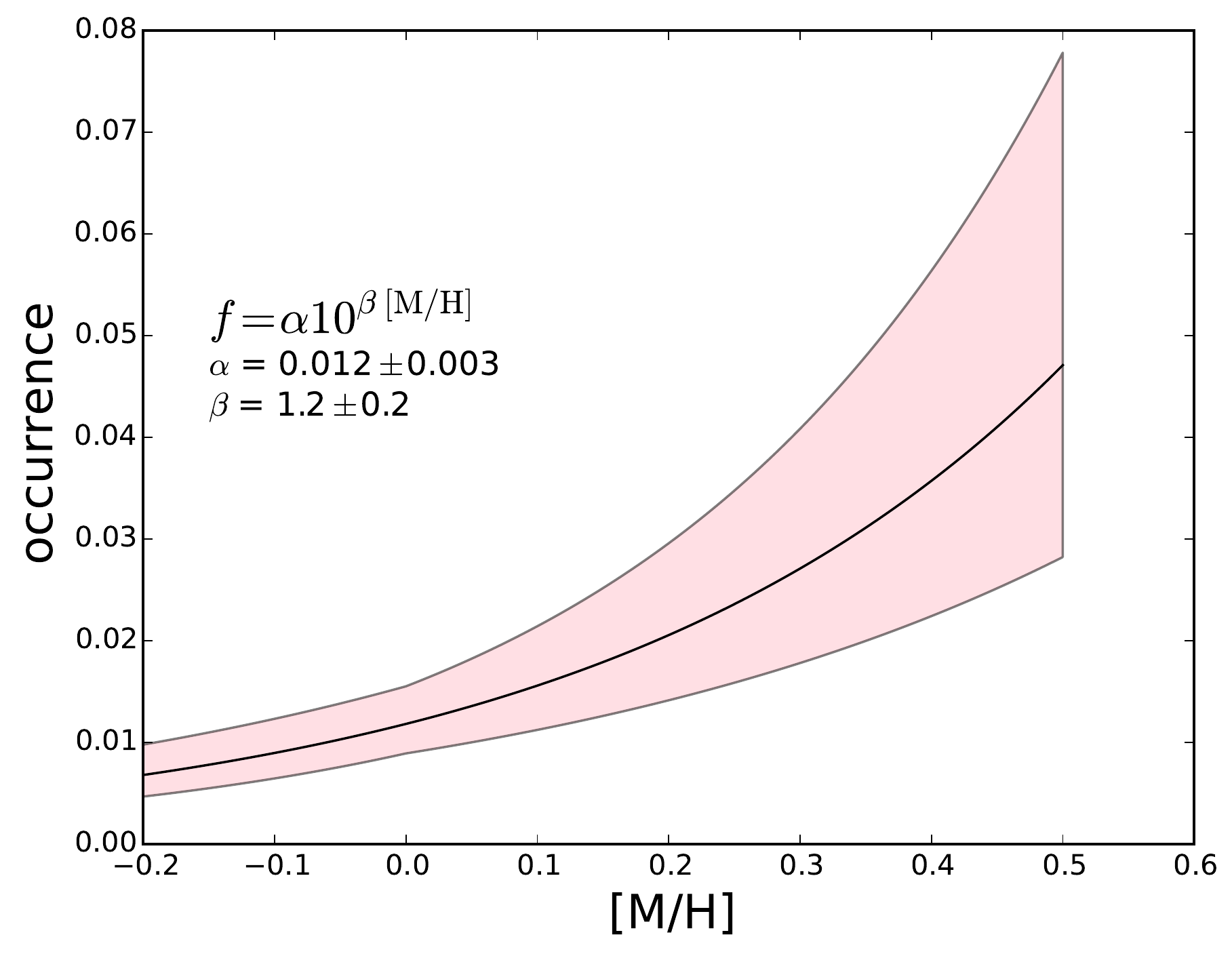}
		}%
        \subfigure
		{%
        \label{fig:second}
       	\includegraphics[width=0.5\textwidth]{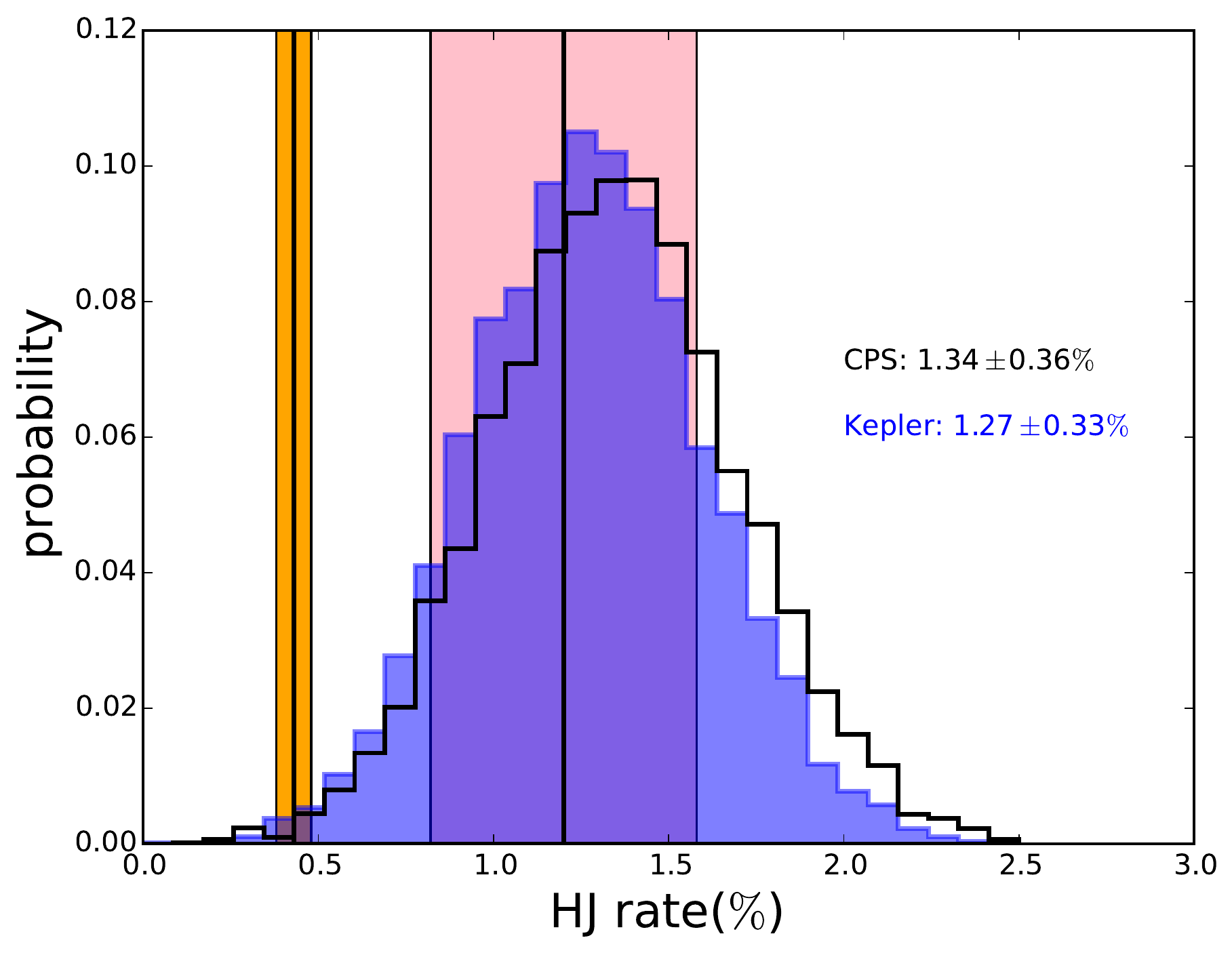}
        } %  ------- End of the first row ----------------------%
\end{center}
\caption{Left panel: the best-fit exponential relation between HJ occurrence probability and host star metallicity [M/H]. Parameter $\beta$ is restrained to a gaussian centered on $\beta = 1.2$ with a 0.2 standard deviation. The pink shade represents 1$\sigma$ uncertainty. Right panel: HJ rate probability distributions calculated according to the best-fit relation and its uncertainty shown in the left panel. The histogram in blue represents the \kep\ sample, and its HJ rate is $1.27\pm 0.33\%$; the histogram outlined in black represents the CPS sample, and its HJ rate is $1.34\pm 0.36\%$. The orange and the pink histograms represents the HJ rate of $0.43\pm 0.05\%$ of the \kep\ field and $1.20\pm 0.38\%$ of solar neighborhood derived in previous observational works.}
\label{fig:HJ_metal_slope12}
\bigskip
\end{figure*}

With the combined sample of 1008 CPS stars, including 13 HJ hosts, we refit the normalization factor $\alpha$ of the relation between the HJ rate and the host star metallicity: $f(\rm HJ) = \alpha 10^{\beta \rm[M/H]}$, with parameter $\beta$ restrained to $\beta = 1.2\pm 0.2$ according to \citetalias{Johnson2010}. To this end, we followed the Bayesian Inference technique described in \citetalias{Johnson2010}. We show the best-fit relation in the left panel of Figure \ref{fig:HJ_metal_slope12}, with the normalization factor $\alpha = 0.012\pm 0.003$. 

We calculated the expected HJ occurrence rates for the \kep\ sample and CPS sample using their metallicity distributions and the best-fit relation. 
For this, we constructed an exponential relation $f(\rm [M/H])$ described by two parameters drawn from Gaussian distributions defined by the best-fit parameter values and their 1$\sigma$ uncertainties. For each star we drew its metallicity [M/H]$_i$ from a Gaussian distribution defined by its measured [M/H] central value and the uncertainty, and then we calculated the total HJ occurrence probability of this sample using $P_{\rm CPS}(\rm HJ) = \left( \sum_{\textit{i}} f(\rm[M/H]_{\textit{i}})\right) / \rm N$, where N is the total number of stars in the sample. We repeated this procedure $2.5\times 10^{6}$ times, each time obtaining the distribution of HJ occurrence rates of the CPS sample. Similarly, we applied the best-fit relation and its uncertainty to our 610 \kep\ dwarf stars, and obtained the HJ occurrence rate distribution of the \kep\ sample. A comparison is shown in the right panel of Figure \ref{fig:HJ_metal_slope12}. The HJ rates of the CPS sample has a mean of 1.34\% with a 0.36\% standard deviation, and the HJ rate of the \kep\ sample has a mean of 1.27\% with a 0.33\% standard deviation. Thus the expected HJ rate of the \kep\ dwarf stars is smaller than that of the solar neighborhood dwarf stars by only around 0.1 percentage point as a result of the metallicity distribution difference and an exponential relation of index $\approx$ 1.2[M/H]. More importantly, this expected HJ rate of the \kep\ sample calculated from its metallicity distribution is still inconsistent with that observed for \kep\ targets.

\subsubsection{Exponential Relation with Relaxed Index}

\begin{figure*}
\begin{center}
%%		\subfigure[caption of first figure]
		\subfigure
		{%
		\label{fig:first}
		\includegraphics[width=0.5\textwidth]{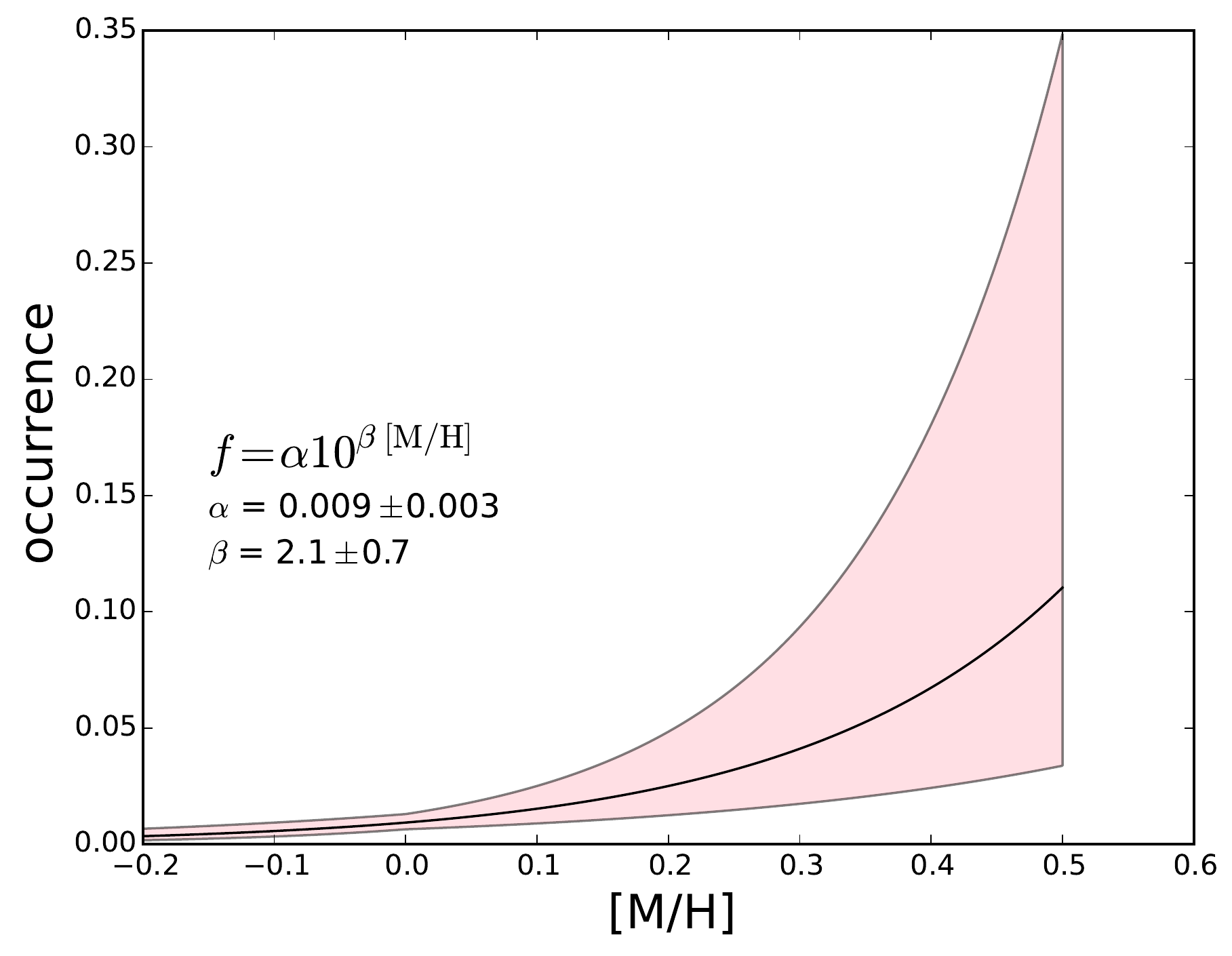}
		}%
        \subfigure
		{%
        \label{fig:second}
       	\includegraphics[width=0.5\textwidth]{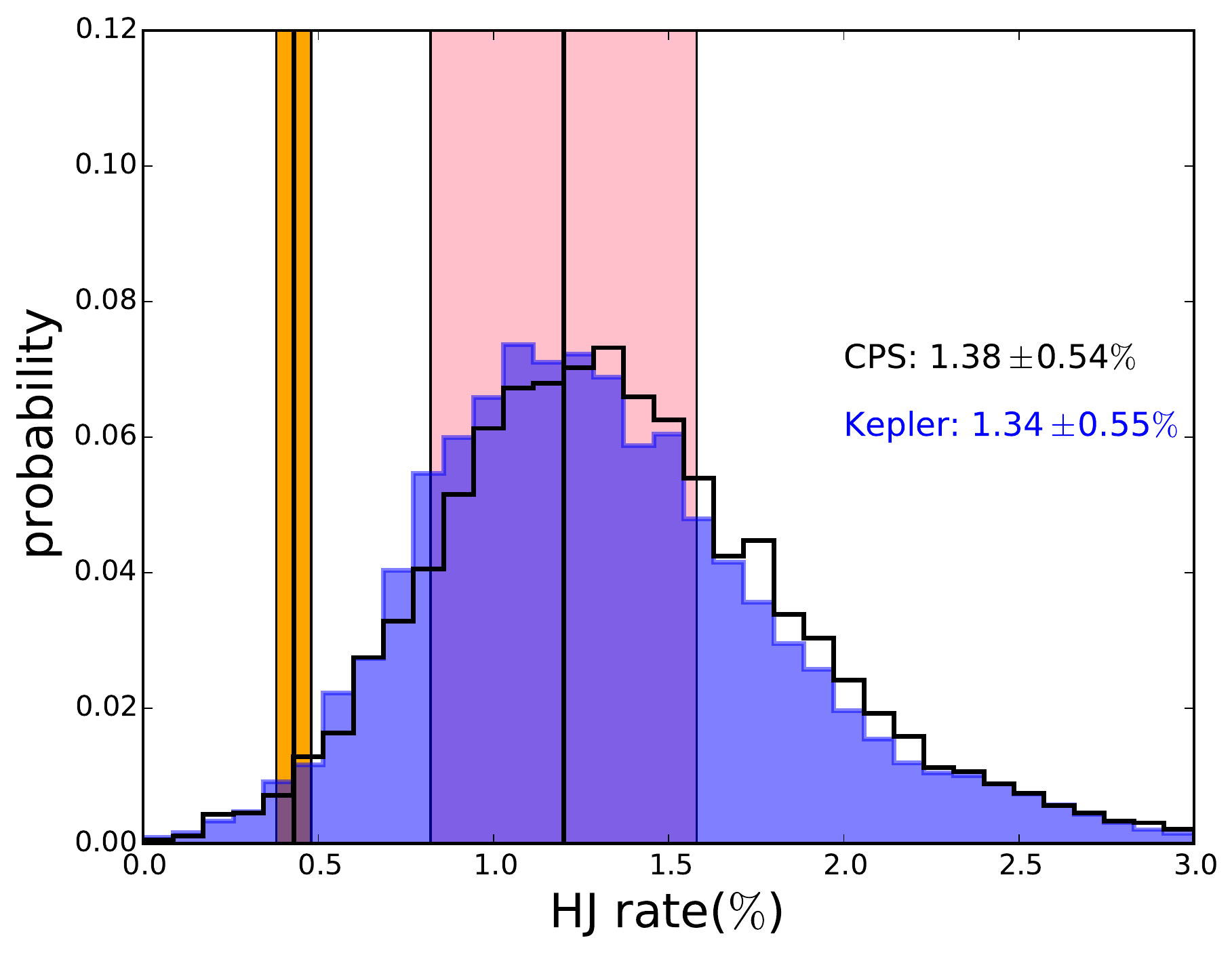}
        } %  ------- End of the first row ----------------------%
\end{center}
\caption{Same as Figure \ref{fig:HJ_metal_slope12} but with parameters $\alpha$ and $\beta$ both relaxed during the fitting. The histogram in blue represents the \kep\ sample, and its HJ rate is $1.34\pm 0.55\%$; the histogram outlined in black represents the CPS sample, and its HJ rate is $1.38\pm 0.54\%$.}
\label{fig:HJ_metal}
\bigskip
\end{figure*}

We relaxed the constraint on the exponential index and refit the relation between the HJ occurrence probability and host star's metallicity. The best-fit relation $f(\rm HJ) = \alpha 10^{\beta \rm[M/H]}$ has $\alpha = 0.009\pm 0.003$ and $\beta = 2.1\pm 0.7$, as is shown in the left panel of Figure \ref{fig:HJ_metal}. Following the procedure described in the $\beta =1.2\pm 0.2$ case, we recalculated the HJ rate probability distributions of the CPS dwarf star sample and the \kep\ dwarf star sample, which are shown in the right panel of Figure \ref{fig:HJ_metal}. The probability distribution of the CPS sample HJ rate has a mean of 1.38\% with a standard deviation of 0.54\%, and the probability distribution of the \kep\ sample HJ rate has a mean of 1.34\% with a standard deviation of 0.55\%. In this case, the probability distributions are broader because of the unrestrained exponential index, and the mean HJ rates of the two samples have even smaller difference: only around 0.04 percentage point.

\bigskip
In summary, according to an exponential relation, the HJ occurrence rates of the \kep\ field and the solar neighborhood should be indistinguishable given measurement uncertainties. The orange and the pink histograms in Figure \ref{fig:HJ_metal_slope12} and Figure \ref{fig:HJ_metal} represent the $0.43\pm 0.05\%$ HJ rate of the \kep\ field derived in \citetalias{Fressin2013} and the $1.20\pm 0.38\%$ HJ rate of the solar neighborhood derived in \citetalias{Wright2012} respectively. The overlaps of the distributions and histograms show that there is only $\lesssim 2\%$ probability that the \kep\ field HJ rate falls into the range $0.43\pm 0.05\%$. We conclude that there must be other factors than metallicity leading to the difference in HJ occurrence.

\subsection{The minimum [M/H] shift required}

\begin{figure}
  \includegraphics[width=\linewidth]{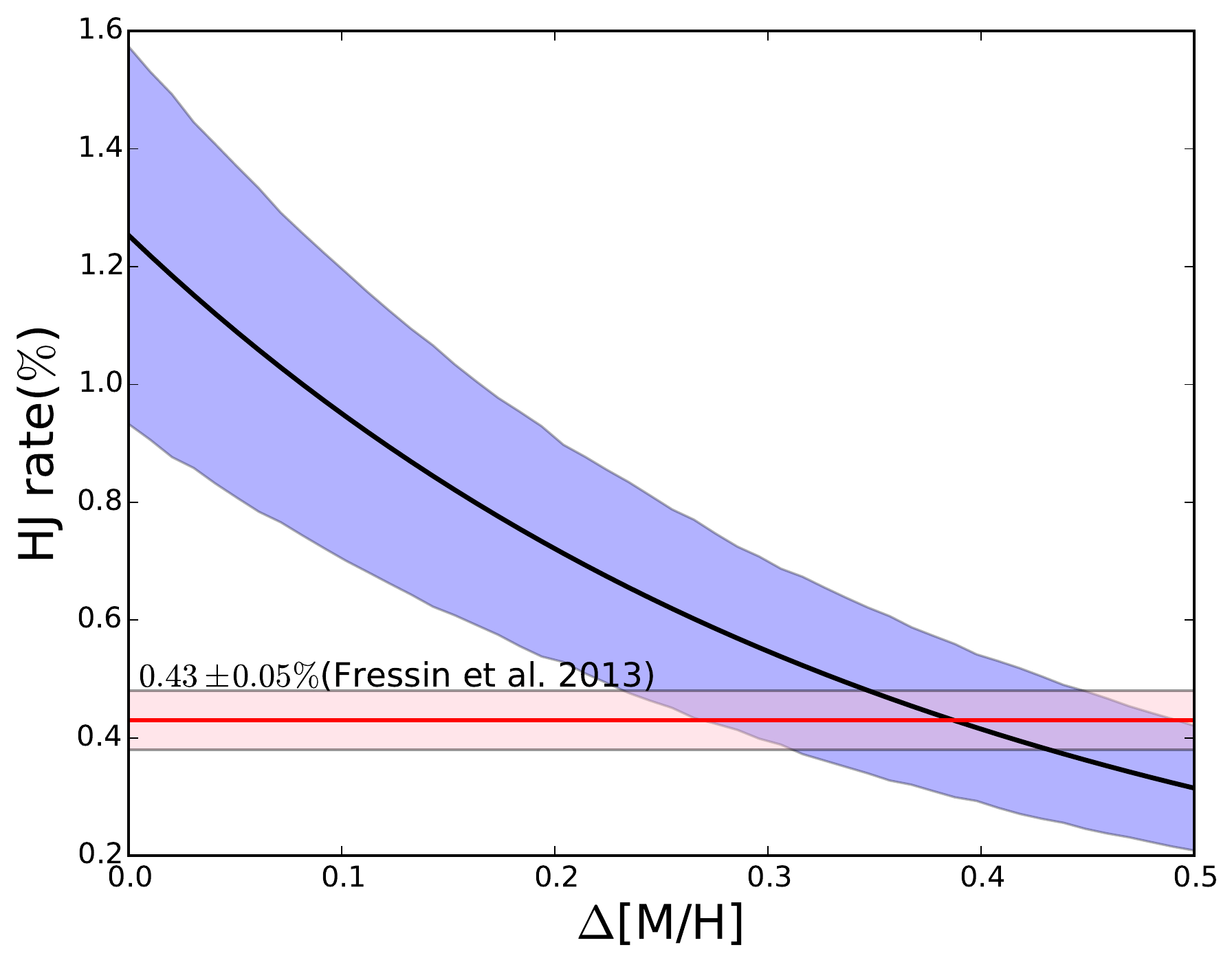}
  \caption{The black curve represents the simulated \kep\ HJ rate as a function of the hypothetical metallcity decrease $\Delta$[M/H], with the blue shade representing the uncertainty. The calculation is performed using the best-fit exponential relation between HJ rate and host star [M/H] with the index fixed to $(1.2\pm 0.2)$[M/H]. The red line and pink shade represents the HJ rate of $0.43\pm 0.05\%$ derived in \citetalias{Fressin2013}. We can see that we need to shift the metallicity distribution by at least 0.3 dex to obtain the $0.43\pm 0.05\%$ HJ rate.}
  \label{fig:shift_slope12}
  \bigskip
\end{figure}

Using the best-fit relation between HJ rate and the host star metallicity (with a restrained index), we investigated how metal-poor the \kep\ field needs to be to perfectly account for the difference in HJ rates. For a certain hypothetical metallicity decrease $\Delta$[M/H], we reduced the metallicity of each star in our sample by $\Delta$[M/H] so that their new metallicities $\rm [M/H]_{new}=[M/H]_{original}-\Delta [M/H]$, and then calculate the new HJ occurrence rate of the sample corresponding to this $\Delta$[M/H] using the best-fit exponential relation between HJ rate and host star metallicity. New HJ rate as a function of the hypothetical metallicity decrease $\Delta$[M/H] is plotted with the black curve in Figure \ref{fig:shift_slope12}, with the blue shade representing the uncertainty. The shaded red horizontal line represents the \kep\ HJ rate of $0.43\pm 0.05\%$ derived in \citetalias{Fressin2013}. We can observe that the [M/H] distribution has to be shifted lower by at least 0.3 dex to perfectly align with the $0.43\%$ HJ rate. Small biases, like those from the $K_P<15$ cut in our Hectochelle sample, are unlikely to be able to account for this difference.

As in the case of restrained exponential index, we measured the HJ rate change as a function of the metallicity decrease $\Delta$[M/H], calculated using the best-fit exponential relation where $\alpha = 0.009\pm 0.003$ and $\beta = 2.1\pm 0.7$. In this case, we found that the [M/H] distribution has to be shifted lower by around 0.2 dex to make the HJ rates match perfectly, similar to our result using the restrained index.

\subsection{Subgiant/Giant Contamination}
The imprecise determination of stellar surface gravities in KIC could lead to imprecise stellar radius and planet radius, thus a different giant planet population. And stellar evolutionary stages may affect the observed HJ rate. The $0.43\%$ HJ occurrence rate measured by \citetalias{Fressin2013} used a dwarf star sample defined by \logg\ $>3.6$. Their \logg\ values were taken from the KIC, which may underestimate the number of evolved stars \citep{Gaidos2013}. According to the criterion used in \citetalias{Fressin2013}, and using \logg\ values from the KIC, we counted that there are 623 stars in our sample with \logg\ $ > 3.6$. \citetalias{Wright2012} HJ occurrence estimate ($1.2\%$) was derived for solar-neighborhood stars using an evolution cut such that only stars deviating upwards from the main-sequence by no more than $\rm \Delta M_V = 2.5$ mag using the main-sequence fit were counted into the background star sample \citepalias{Wright2012}. This criterion corresponds roughly to \logg\ $ > 3.5$ \citep{Wang2015}. There are 610 stars in our sample with Hectochelle spectroscopic \logg\ matching this limit. From this we estimated that only $\simeq2\%$ of the stars in the \citetalias{Fressin2013} sample are subgiants/giants misidentified as dwarf stars. This low rate of subgiant/giant contamination has a negligible effect on the HJ occurrence comparison.

\subsection{Other Reasons of the HJ Occurrence Rate Difference}

\citet{Wang2015} argued that the CPS survey targets are biased towards single stars, so \kep\ sample stellar multiplicity should be higher than the CPS stellar multiplicity, and this could result in a lower planet occurrence rate of \kep\ sample because planet formation is suppressed in multiple star systems \citep{Kraus2016, Wang2015}. According to the statistics in \citet{Kraus2016}, in a volume limited sample, about 20\% of all solar-type stars are disallowed from hosting planetary systems due to the presence of a binary companion, while in a flux limited sample, the ratio should be even higher due to Malmquist bias. We estimate the planetary formation suppression level of the Kepler flux limited sample as follows. About 1/2 of all stars are binaries \citep{Raghavan2010}, of which 1/2 are bright enough to be seen even if they reside outside of the nominal volume. Therefore $(1/4)/(1+1/4)=1/5$ of all stars are binaries outside of the nominal volume. Of those binaries, around 1/2 have semi-major axis smaller than 50 AU \citep{Raghavan2010}, therefore they have a 2/3 probability of being disallowed from hosting planets according to \citet{Kraus2016}. So binaries outside of the nominal volume contribute $1/5\times 1/2\times 2/3 \approx 7\%$ planet suppression rate, while all stars in the nominal volume contribute $4/5\times 20\% = 16\%$ suppresion rate. As a result, $7\%+16\% = 23\%$ of all stars in a flux limited sample are disallowed from hosting planets due to binary companions. However, \citet{Santerne2016} used the occurrence rate of eclipsing binaries with transit depth deeper than 3\% as a proxy of the stellar multiplicity rate, and found that the CoRoT field has an eclipsing binary rate of $0.94\pm 0.02\%$, higher than the $0.79\pm 0.02\%$ eclipsing binary rate of the \kep\ field. The fact that we didn't observe an even lower HJ rate in the CoRoT field casts doubt on the multiplicity as a primary factor. In addition, although planet formation is suppressed in general in multiple star systems, it is not clear whether gas giant formation is suppressed more or less efficiently than small planet formation. Therefore the HJ deficiency level in the \kep\ field as a result of the \kep\ star multiplicity rate is unclear.

Transit surveys have detected HJs around A-type stars ($M_* \gtrsim 1.4~M_\odot$) and a few around M dwarfs, but the studies of HJ occurrence rates using RV samples have been limited to the mass range $0.8-1.2M_{\odot}$. While the occurrence rates of short-period, giants planets have been limited to a narrow stellar mass range, \citet{Johnson2010} studied the occurrence rate of giant planets out to 2.5 AU around stars ranging from $\sim 0.2 M_\odot$ to $2.0 M_\odot$, and derived an empirical relation between the occurrence rate and the host star mass: $f(M_*)\propto (M_*/M_{\odot})^{1.0\pm 0.3}$. Thus, the occurrence rate of giant planets increases with increasing host star mass. Even if the HJ occurrence rate follows this same stellar mass dependence, we don't expect stellar mass to impact our study since all stars in our \kep\ sample and the CPS sample are intermediate mass FGK stars in the mass range $0.8-1.2~M_\odot$. 

\bigskip

\section{Summary} \label{section6}

We calibrated Hectochelle spectra for spectroscopic analysis by developing a functional form to emulate the Hectochelle continuum profile. This was achieved by fitting the Hectochelle twilight spectra with the normalized high-resolution NSO solar spectrum multiplied with our test continuum functional forms. From there we combined the continuum function with the calibrated Kurucz library to measure stellar parameters with Hectochelle spectra by searching for the minimum of best-fit figure-of-merits on the grid points in the 3D space of (\teff, \logg, [M/H]). 

To test the reliability of our analysis and estimate errors on our parameters we performed three empirical tests: 1) we analyzed 203 twilight (solar) spectra from Hectochelle, 2) we compared our parameters of 28 stars with measurements in literatures or derived from SPC and higher quality data, and 3) analyzed 36 stars in the open cluster NGC 752. Based on these comparisons we estimated our errors to be 100 K for \teff, 0.1 dex for [M/H] and 0.1 for \logg.

We applied our method to 776 stellar spectra in our \kep\ sample to derive \teff, [M/H], and \logg.  Our \teff\ values agree well with those from KSPA \citep{Huber2014q17}, except for a systematic disagreement at \teff$>5500$ K, which we attribute to a known difference between photometric and spectroscopic \teff\ values in this range \citep[also seen by][]{Santerne2016}. Comparison of \logg\ shows larger scatter, especially for evolved stars, which we mostly attribute to unreliable \logg\ determinations from photometry \citep[see][]{Gaidos2013}. 

Taking only dwarf stars with \logg\ $>3.5$ in our sample, we obtained a sub-sample of 610 stars, and presented their metallicity distribution in Figure \ref{fig:stat_Z}. We calculated that the mean of our \kep\ dwarf star metallicity distribution is $\rm [M/H]_{dwarf}=-0.045\pm 0.009$ dex and the standard deviation of the distribution is 0.225 dex. The mean value of our \kep\ dwarf star metallicity distribution agrees with that of the metallicity distribution of the LAMOST 14000 \kep\ dwarf star sample, and the two-sample KS test shows that the two distributions are highly consistent. In addition, \citet{Santerne2016} investigated the metallicities of 37 \kep\ stars with high-resolution spectroscopy and reported sub-solar average value, in agreement with our result. 

We compared the metallicity distribution of our \kep\ sample with that of the CPS sample. To this end we combined previous metallicity measurements from \citetalias{Brewer2016} and \citet{Valenti2005}, and derived and updated CPS sample metallicity distribution, with a mean of $\rm [M/H]_{CPS} = -0.005 \pm 0.006$ and a standard deviation of $\sigma_{\rm{[M/H]}}=0.187$ dex. A two-sample KS test gives a p-value of $7.2\times 10^{-6}$, indicating the \kep\ [M/H] sample and the CPS [M/H] sample are distinct.

To estimate if this metallicity difference is sufficient to explain the HJ occurrence rate differences we refit the HJ rate-metallicity exponential relation using the CPS sample of 1008 stars for two cases: the exponential index restrained and relaxed. We applied the best-fit relations on the [M/H] distributions of the two samples to calculate their expected HJ occurrence rate. The results show that the HJ rates of the \kep\ field and the solar neighborhood could only be different by $\lesssim 0.1$ percentage point. And using the best-fit HJ rate-metallicity relations, we find that the Kepler [M/H] distribution has to be shifted lower by at least 0.2 dex to match the observed Kepler HJ rate. We conclude that the $0.43\pm 0.05\%$ \kep\ field HJ rate and the $1.20\pm 0.38\%$ solar neighborhood HJ rate cannot be reconciled by metallicity differences alone. In addition, we checked the subgiant/giant contamination of the \kep\ star sample, finding that only $\lesssim 2\%$ subgiants/giants were misidentified as dwarf stars, much smaller than the required contamination rate to explain the HJ rate differences.

We also discussed other possible reasons responsible for the HJ rate discrepancy. The high multiplicity of Kepler targets, the imprecise determination of Kepler stellar surface gravities, the mass distribution discrepancy between the Kepler sample and the RV sample, and the inaccurate false positive rates are all potential reasons. 

While no individual reason can explain the observed HJ rate discrepancy, it could be a combination of all these factors. The effect of each factor is not known with sufficient precision to say with certainty if the combination is sufficient, or we need to explore other explanations. However, when Gaia parallaxes become available for the entire Kepler sample, we will have more precise measurements of stellar properties like masses and radius as well as a better estimation of the multiplicity of the \kep\ field, hence providing a more reliable estimation of the relative contribution of these factors.

\bigskip

\section{Acknowledgements}

We would like to thank Professor Matthew Walker and Professor Edward Olszewski for the help with Hectochelle observations. We also thank Doctor Nelson Caldwell for compiling the Hectochelle twilight spectra that we used in Section \ref{section4}, and thank Doctor John Brewer for useful discussions about the trend in spectroscopic metallicity as a function of \teff.

This research has made use of the NASA Exoplanet Archive, which is operated by the California Institute of Technology, under contract with the National Aeronautics and Space Administration under the Exoplanet Exploration Program.

Some of the data presented in this paper were obtained from the Mikulski Archive for Space Telescopes (MAST). STScI is operated by the Association of Universities for Research in Astronomy, Inc., under NASA contract NAS5-26555. Support for MAST for non-HST data is provided by the NASA Office of Space Science via grant NNX09AF08G and by other grants and contracts. 

Jason L.Curtis is supported by the National Science Foundation Astronomy and Astrophysics Postdoctoral Fellowship under award AST-1602662 and NASA grant NNX16AE64G.

\bigskip

\bibliographystyle{apa}
\bibliography{mybib}

\clearpage
%\pagebreak

%\vspace*{1\baselineskip}

\renewcommand*{\arraystretch}{1.4}
\begin{table*}
\caption{Stellar Parameters of 776 \kep\ stars.}
%\centering
\begin{tabularx}{\textwidth}{nnnnnnnn}%{  c {\textwidth/7} c {\textwidth/7} c {\textwidth/7} c {\textwidth/7} c {\textwidth/7} c {\textwidth/7} c {\textwidth/7} c }
%\tableline
\hline\hline
KeplerID & RA(J2000) & Dec(J2000) & $K_p$(mag) & \teff (K) & \logg ($\rm cm/s^2$) & [M/H] & Binary \\
\hline
6766793 & 19h13m30.6s & +42d17m35.3s & 14.72 & 5830 & 3.90 & -0.08 & N \\ 
6766990 & 19h13m49.1s & +42d13m20.0s & 14.84 & 4680 & 4.48 & 0.16 & N \\ 
6767100 & 19h13m56.5s & +42d14m3.1s & 14.60 & 5580 & 4.77 & -0.13 & N \\ 
6767489 & 19h14m33.9s & +42d16m55.2s & 12.34 & 4930 & 3.45 & 0.12 & N \\ 
6767829 & 19h15m4.5s & +42d14m48.5s & 13.03 & 4820 & 2.95 & -0.30 & N \\ 
6851516 & 19h12m25.1s & +42d21m40.1s & 14.88 & 5480 & 4.09 & 0.11 & N \\ 
6851792 & 19h12m50.6s & +42d19m24.0s & 13.48 & 6100 & 4.11 & -0.04 & N \\ 
6851944 & 19h13m4.5s & +42d23m33.8s & 14.76 & 4820 & 4.43 & 0.11 & P \\ 
6852013 & 19h13m10.0s & +42d23m56.7s & 13.98 & 5960 & 3.66 & 0.08 & N \\ 
6852189 & 19h13m25.6s & +42d22m33.8s & 14.38 & 6010 & 3.90 & -0.19 & N \\
%\onecolumn
\hline
\end{tabularx}
\begin{tablenotes}
\small
\item \textbf{NOTE.} 
\item 1. In the "Binary" column, "N" represents "not a binary" and "P" represents "potential binary".
\item 2. All stars have uniform empirical uncertainties on \teff, \logg\ and [M/H]: $\sigma_{\Teff}=100$ K, $\sigma_{\Logg}=0.1$ and $\sigma_{\MH}=0.1$. The uncertainties are estimated in section \ref{section4} by picking out stars with multiple measurements and comparing the best-fit parameters for each observation of the same stars.
\item (A portion of this table is shown for form and content. The full table will be available online.)
\end{tablenotes}
\label{tbl-1}
\end{table*}
%\tablenotetext{1}{A portion of this table is shown for form and content. The full table will be available online.}
%\tablecomments{\footnotesize In the "Binary" column, "N" represents "not a binary" and "P" represents "potential binary".}

%\end{deluxetable}

\end{document}